\documentclass[preprint,notoc]{JHEP3}
\usepackage{epsfig}
\def\eslt{\not\!\!{E_T}}
\def\mslash{\not\!\!{m}}
\def\to{\rightarrow}

\def\bi{\begin{itemize}}
\def\ei{\end{itemize}}
\def\be{\begin{equation}}
\def\ee{\end{equation}}
\def\te{\tilde e}

\def\tu{\tilde u}

\def\tst{\tilde t}
\def\ttau{\tilde \tau}
\def\tmu{\tilde \mu}
\def\tg{\tilde g}

\def\tell{\tilde\ell}

\def\tw{\widetilde W}
\def\tz{\widetilde Z}
\def\alt{\stackrel{<}{\sim}}

\title{
Linear Collider Capabilities for Supersymmetry \\
in Dark Matter Allowed Regions\\ 
of the mSUGRA Model
}

\author{Howard Baer, Alexander Belyaev and Tadas Krupovnickas
\\ Department of Physics, Florida State University\\ 
Tallahassee, FL 32306, USA\\
E-mail: \email{baer@hep.fsu.edu}, \email{belyaev@hep.fsu.edu}, 
\email{tadas@hep.fsu.edu}}
\author{Xerxes Tata
\\ Department of Physics and Astronomy, University of Hawaii,\\
Honolulu, HI 96822, USA \\ 
E-mail: \email{tata@phys.hawaii.edu}}

\preprint{\vbox{\hbox{FSU-HEP-031104} \vspace{0.2cm}
                \hbox{UH-511-1040-03}}} 

\abstract{
Recent comparisons of minimal supergravity (mSUGRA) model predictions with
WMAP measurements of the neutralino relic density
point to preferred regions
of model parameter space.
We investigate the reach of linear colliders (LC) with
$\sqrt{s}=0.5$ and 1~TeV for SUSY in the 
framework of the mSUGRA model.
We find that LCs can cover the entire stau co-annihilation 
region provided $\tan\beta \alt 30$.
In the hyperbolic branch/focus point (HB/FP) region of parameter space,
specialized cuts are suggested to increase the reach in this important
``dark matter allowed'' area. 
In the case of the HB/FP region, the reach of a LC 
extends well past the reach of the CERN LHC.
We examine a case study in the HB/FP region, and show that
the MSSM parameters $\mu$ and $M_2$ can be sufficiently well-measured
 to demonstrate that one would indeed be in the HB/FP region, 
where the lightest chargino and neutralino have a substantial
higgsino component.
}


\begin{document}

\section{Introduction}
\label{sec:intro}

In recent years, supersymmetric models have become increasingly constrained
by a variety of measurements~\cite{constraints}. 
These include determination of the
branching fraction $BF(b\to s\gamma )$~\cite{bsg}, the muon anomalous
magnetic moment $a_\mu =(g-2)_\mu/2$ \cite{e821} and most recently, 
the tight restriction on the density of relic dark matter from the Big Bang, 
as determined by the WMAP experiment\cite{wmap}.
Analysis of WMAP and other data sets have determined a preferred range for
the abundance of cold dark matter\cite{wmap,dimitri}:
\be
\Omega_{CDM}h^2=0.1126^{+0.0161}_{-0.0181},\ \ \ \ 2\sigma\ {\rm level}.
\label{wmaprange}
\ee  

Within the minimal supergravity (mSUGRA) framework\cite{msugra}, the
lightest neutralino is usually the lightest SUSY particle. Since
$R$-parity is assumed to be conserved, this neutralino is stable and
provides a good candidate for cold dark matter\cite{haim}. While it is
possible that the relic density of neutralinos may make up almost all the
cosmological dark matter, the possibility that dark matter, like visible
matter, is made up of several components cannot be excluded at this point.
In our analysis we will, therefore, interpret the WMAP measurement
(\ref{wmaprange}) as an {\it upper} bound, $$\Omega_{\tz_1}h^2 < 0.129,$$
on the relic density in neutralinos.
The mSUGRA model is characterized by four SUSY
parameters together with a sign choice,
\be
m_0,\ m_{1/2},\ A_0,\ \tan\beta\ \ {\rm and}\ sign(\mu ).
\ee
Here $m_0$ is the common mass
of all scalar particles at $M_{GUT}$, $m_{1/2}$ is the common 
gaugino mass at $M_{GUT}$, $A_0$ is the common trilinear soft term at
$M_{GUT}$, $\tan\beta$ is the ratio of Higgs field vacuum expectation 
values at the scale $M_Z$, and
finally the magnitude -- but not the sign -- of the superpotential
$\mu$ term is determined by the requirement of radiative electroweak 
symmetry breaking (REWSB). In addition, we take $m_t=175$~GeV.

The recent measurements of $\Omega_{CDM}h^2$, 
$BF(b\to s\gamma )$ and $a_{\mu}$ 
have considerably modified our expectations for the regions of
mSUGRA model parameter space that may be realized 
in nature. 
Several years ago, relatively low values of $m_0$ and $m_{1/2}$
appeared to be 
preferred because these led to sufficiently light sleptons which
in turn led to efficient neutralino annihilation in the early universe 
via $t$-channel sfermion exchange diagrams; this
resulted in values of $\Omega_{CDM} h^2<1$, and a universe
at least as old as its oldest constituents\cite{gjk,bb}.
The low $m_0$ and $m_{1/2}$ region of parameter space has been dubbed
by Ellis {\it et al.} as the ``bulk'' annihilation region\cite{ellis}. 
The bulk annihilation region was also favored by fine-tuning 
estimates in the mSUGRA model\cite{diego}.

Currently, the bulk annihilation region of the mSUGRA model is disfavored by:
\begin{enumerate}
\item a value of $m_h$ typically below bounds from LEP2, which require
$m_h>114.4$~GeV in the case of a SM-like light Higgs scalar
$h$\cite{lep2h};
\item a value of $BF(b\to s\gamma )$ that is either above ($\mu <0$) or
below ($\mu >0$) current measurements\footnote{A combination of
measurements from the ALEPH, BELLE and CLEO experiments yield $BF(b\to
s\gamma )= (3.25\pm 0.54)\times 10^{-4}$, while the SM prediction is
$(3.6\pm 0.3)\times 10^{-4}$~\cite{bsg}.};
\item for larger values of $\tan\beta$, 
the bulk region leads to large negative ($\mu <0$) or large 
positive ($\mu >0$) contributions to the muon magnetic dipole moment.
Because of uncertainty
in the SM prediction of $a_{\mu}$,
caution is advised in the interpretation of what the E821 data tell us
about the existence of physics beyond the SM.\footnote{
In a recent analysis, Davier {\it et al.}\cite{davier} find 
$\Delta a_\mu =(22.1\pm 11.3)\times 10^{-10}$
($(7.4\pm 10.5)\times 10^{-10}$) [errors added in quadrature] 
depending on whether the hadronic vacuum polarization is
estimated using $e^+e^-\to hadrons$ ($\tau$ decay) data.}
\end{enumerate}

The favored regions of mSUGRA parameter space now 
include\cite{wmap_pap,sug_chi2,ellis_likely}
\begin{itemize}
\item the stau co-annihilation region at low $m_0$ 
where $m_{\ttau_1}\simeq m_{\tz_1}$, and where
$\ttau_1 -\tz_1$ and $\ttau_1-\bar{\ttau}_1$ 
annihilation in the early universe also serve to 
reduce the neutralino relic density to sufficiently low values\cite{stau},
\item the $A$-annihilation funnel at large $\tan\beta$ where
$m_H$ and $m_A\simeq 2m_{\tz_1}$ and $\tz_1\tz_1\to A,\ H\to f\bar{f}$ 
($f$'s are
SM fermions) through the very broad $A$ and $H$ 
resonances\cite{Afunnel}, and
\item the hyperbolic branch/focus point region\cite{ccn,fmm,bcpt} (HB/FP) at
large $m_0$ near the edge of parameter space where $\mu$ becomes
small, and the $\tz_1$ has a significant higgsino component which
facilitates a large annihilation rate\cite{bb2,fmw,bb}. The location of
this region is very sensitive to the value of $m_t$ \cite{bktsens}.
\end{itemize}
The HB/FP region predicts multi-TeV squark and slepton masses so
that supersymmetric contributions to $BF(b\to s\gamma )$ and $a_\mu$ 
are suppressed; these measurements are thus expected to be close
to the SM predictions.
Furthermore, the relatively high scalar masses help suppress 
potential flavor changing and $CP$-violating supersymmetric processes,
and offer at least a partial decoupling solution to the SUSY flavor and $CP$
problems\cite{fmm}. 
Finally, Feng {\it et al.} have shown that 
the low $m_{1/2}$ part of the hyperbolic branch has relatively 
low fine-tuning\cite{fmm}, in spite of $m_0$ being very large. 

Given these expectations for where supersymmetry might lie, it makes
sense to re-evaluate the prospects for SUSY searches at various collider
and dark matter search experiments. In Ref. \cite{bkt}, the reach of 
the Fermilab Tevatron for isolated trilepton events from $\tw_1\tz_2$ 
production was extended to very large $m_0$ values to
include the HB/FP region. In the HB/FP region, production cross sections 
increase due to decreased chargino and neutralino masses, but the visible
energy from $\tw_1\to f\bar{f}'\tz_1$ and $\tz_2\to f\bar{f}\tz_1$
decays also decreases, reducing detection efficiency. It was 
shown that the Tevatron reach in $m_{1/2}$ for isolated $3\ell$ events did in
fact {\it increase} in the HB/FP region. 
The reach of the CERN LHC was also
worked out in Ref. \cite{bbbkt}, where it was found that 
values of $m_{1/2}\sim 1400$~GeV ($\sim 700$~GeV) 
could be probed in the stau co-annihilation region 
(HB/FP region) with an integrated luminosity of  100 fb$^{-1}$. 
For higher values of $m_{1/2}$ in the
HB/FP region, $m_{\tg}$ becomes very large while the visible energy from 
$\tw_1$ and $\tz_2$ decays becomes small,  so that signal detection 
becomes difficult. Thus, the high $m_{1/2}$ part of the hyperbolic branch
currently seems beyond LHC reach.
It is interesting to note that in the HB/FP region of the mSUGRA
model, the higgsino component of the $\tz_1$ is sufficiently large to
yield direct neutralino-nucleus scattering rates within the
reach of Stage 3 direct dark matter detection experiments such as
Genius, Cryoarray and Zeplin-4\cite{ddmsearch}.
In addition, in the HB/FP region one may expect detectable rates for
detection of neutrinos arising from neutralino annihilation in the 
core of the sun or the earth, and also large rates for cosmic
photons, positrons and antiprotons due to DM annihilation 
in the galactic halo\cite{fmw,deboer}.

In the HB/FP region, since $|\mu|$ becomes small, 
charginos are light. This then implies that there 
would be a large rate for chargino pair production at 
$e^+e^-$ linear colliders (LCs) operating with center-of-mass energy
$\sqrt{s}\simeq 0.5-1$ TeV over
most of the HB/FP region. However, since the $\tw_1 -\tz_1$ mass
gap also becomes small, it is not clear that linear collider
experiments would access the entire kinematically allowed chargino pair
production region. 

Our main goal in this paper is to assess the reach of linear colliders
for SUSY in the mSUGRA model\cite{bmt,nlcreach}, 
paying particular attention to the HB/FP
region. We note here that previous reach estimates of linear colliders
for SUSY in the mSUGRA model extended only up to $m_0$ values as high as
800~GeV\cite{bmt}-- well below the HB/FP region. 
We find that a linear collider, using
standard cuts for chargino pair events, can explore the low $m_{1/2}$ 
portion of the HB/FP region. 
To explore the high $m_{1/2}$ part of the 
hyperbolic branch, new specialized cuts are suggested. With these
cuts, it appears possible to probe
essentially all of the HB/FP region (up to $m_{1/2}= 1.6$~TeV)
where charginos satisfy the LEP2 bounds,
and where chargino pairs are 
kinematically accessible
at a linear collider (LC).
This then provides the first example of a SUSY parameter space region 
which is accessible to linear $e^+e^-$ colliders, while likely 
remaining out of reach of LHC experiments! This is especially interesting
since the HB/FP region is one of the three qualitatively
different  mSUGRA parameter space regions allowed 
by dark matter and other constraints.

If indeed a SUSY signal from charginos is detected, the next step would be 
to try and determine the associated weak scale parameters:
$\mu$, $M_2$ and $\tan\beta$\cite{jlc1}. 
We explore a particular 
case study in the low $m_{1/2}$ region of the HB/FP region, and show that
at least in this case $\mu$ and $M_2$ should be measurable. This measurement
would give a firm indication of the large higgsino content of the 
light chargino and $\tz_1$ and, together with the 
fact that sfermions are not detected either at the LC or at the LHC, 
provide a strong indication that SUSY in fact 
lies in the HB/FP region.

The remainder of this paper is organized as follows. In Sec.~2, 
we examine the problems associated with
detecting SUSY in mSUGRA parameter space at an $e^+e^-$ LC, 
with special emphasis on the 
HB/FP region, and show that these can be overcome. 
We propose novel cuts that allow the large
$m_{1/2}$ part of the hyperbolic branch to be explored via chargino pair
production. We then show our projections for the SUSY 
reach of a LC operating at $\sqrt{s}=0.5$~TeV or $\sqrt{s}=1$~TeV,
assuming 100 fb$^{-1}$ of integrated luminosity.
In Sec.~3, we compare the reach of a LC
with the reach of Fermilab Tevatron luminosity upgrades
and of the CERN LHC. Furthermore, we
also show the dark matter allowed region of
mSUGRA parameter space to elucidate the performance of these experiments
in this preferred region of parameter space.
In Sec.~4, we perform a case study in the HB/FP region, 
where chargino pair production is accessible to a $\sqrt{s}=0.5$ TeV LC.
The masses $m_{\tw_1}$ and $m_{\tz_1}$ can be measured via 
energy distribution end-points. We find that these measurements,
along with measurement of the total chargino pair cross section, allow 
a determination of the underlying SUSY parameters $\mu$ and $M_2$,
although $\tan\beta$ remains relatively undetermined. 
In Sec.~5, we present our conclusions.

\section{Reach of a Linear Collider in the mSUGRA model}

In our signal and background computations, we use ISAJET
7.69\cite{isajet} which allows for the use of polarized beams, and also
allows for convolution of subprocess cross sections with electron parton
distribution functions (PDFs) arising from both initial state
bremsstrahlung and also beamstrahlung\cite{PChen}.  We use the ISAJET
toy detector CALSIM with calorimetry covering the regions $-4<\eta <4$
with cell size $\Delta\eta\times\Delta\phi = 0.05\times
0.05$. Electromagnetic energy resolution is given by $\Delta
E_{em}/E_{em}=0.15/\sqrt{E_{em}}\oplus 0.01$, while hadronic resolution
is given by $\Delta E_h/E_h=0.5\sqrt{E_h}\oplus 0.02$, where $\oplus$
denotes addition in quadrature.  Jets are
identified using the ISAJET jet finding algorithm GETJET using a fixed
cone size of $\Delta R=\sqrt{\Delta\eta^2+\Delta\phi^2}=0.6$, modified
to cluster on energy rather than transverse energy.  Clusters with
$E>5$~GeV and $|\eta (jet)|<2.5$ are labeled as jets.  Muons and
electrons are classified as isolated if they have $E>5$~GeV, $|\eta_\ell
|<2.5$, and the visible activity within a cone of $R=0.5$ about the
lepton direction is less than $max(E_\ell/10\ {\rm~GeV},1\
{\rm~GeV})$. Finally, jets originating from $b$-quarks are tagged as
$b$-jets with an efficiency of 50\%.
\subsection{Review of previous reach assessment} 

The reach of a $\sqrt{s}=0.5$ TeV LC for supersymmetry 
has previously been evaluated in
Ref.~\cite{bmt} assuming an integrated luminosity of
20 fb$^{-1}$. Reach contours were presented for the case 
of the mSUGRA model in the $m_0\ vs.\ m_{1/2}$
plane for $A_0=0$, $\tan\beta =2$ and 10, and $\mu {>\atop <}0$. 
The reach plots in that study were limited to $m_{1/2}<600$~GeV and
$m_0<800$~GeV, {\it i.e.} well outside the HB/FP region. 

The region of the $m_0-m_{1/2}$ plane where there should be an
observable SUSY signal in LC experiments 
consists of three distinct pieces.
\bi
\item At low $m_0$ with $m_{1/2}\sim 300-500$~GeV, slepton pair
production occurs at large rates. The signal is a pair of opposite
sign/same flavor leptons plus missing energy.  Tsukamoto {\it et
al.}\cite{jlc1} suggested cuts of {\it i}) 5~GeV $<E(\ell )<$ 200~GeV,
{\it ii}) 20~GeV $<E_{vis.}<\sqrt{s}-100$~GeV, {\it iii})
$|m(\ell\bar{\ell})-M_Z|>10$ GeV, {\it iv}) $|\cos\theta (\ell^\pm
)|<0.9$, {\it v}) $-Q_\ell\cos\theta_\ell <0.75$, {\it vi})
$\theta_{acop}>30^\circ$, {\it vii}) $E_T^{mis}>25$~GeV and {\it viii})
veto events with any jet activity.  Here, $\theta_{acop} \equiv \pi -
\cos^{-1}(\hat{p}^+_x \hat{p}^-_x + \hat{p}^+_y \hat{p}^-_y)$.  The
reach was evaluated by running with right-polarized electron beams where
$P_L(e^- )=-0.9$.  The beam polarization maximizes
$\tell_R\bar{\tell}_R$ pair production, while minimizing background from
$W^+W^-$ production. 

\item At low $m_{1/2}$ values, chargino pair production occurs at a large
rate. To search for chargino pairs, one may look for 
$1\ell +2j +E^{mis}$ events. Following Ref. \cite{jlc1}, 
it was required in Ref.~\cite{bmt} 
to have one isolated lepton plus two jets with
{\it i}) 20~GeV $<E_{vis}<\sqrt{s}-100$~GeV, {\it ii}) if $E_{jj}>200$~GeV, 
then $m(jj)<68$~GeV, {\it iii}) $E_T^{mis}>25$~GeV, {\it iv}) 
$|m(\ell \nu )-M_W|>10$~GeV for a $W$ pair hypothesis, 
{\it v}) $|\cos\theta(j)|<0.9$, $|\cos\theta (\ell )|<0.9$,
$-Q_\ell\cos\theta_\ell <0.75$ and $Q_\ell\cos\theta (jj)<0.75$, 
{\it vi}) $\theta_{acop}(WW)>30^\circ$ for a $W$ pair hypothesis.
The reach for $1\ell +2j +\eslt $  events from chargino pair
production was evaluated using a left polarized beam with 
$P_L=+0.9$.
\item Finally, there exists a small region around $m_0\sim 200-500$~GeV
and $m_{1/2}\sim 300-350$~GeV where neither slepton pairs nor
chargino pairs are kinematically accessible, but where $e^+e^-\to\tz_1\tz_2$
is. In this case, the decay $\tz_2\to\tz_1 h$ was usually
found to be dominant. Since $h\to b\bar{b}$ with a large branching fraction,
$b\bar{b}+\eslt $ events were searched 
for with two tagged $b$-jets, $E_T^{mis}>25$~GeV and 
$30^\circ <\Delta\phi_{b\bar b}<150^\circ $. Imposing a missing mass
cut $\mslash>340$~GeV eliminated almost all SM backgrounds, so that
a signal cross section of 10 fb would yield 10 signal events for
integrated luminosity 20~fb$^{-1}$, where the signal efficiency was found to 
be 6\%. Beam polarization of $P_L=+0.9$ was used.
\ei
In the following, we refer to these as the ``standard cuts''.

The ultimate reach contours found in Ref. \cite{bmt} generally track
the boundary of the kinematically 
allowed regions for $\tw_1$ and $\te_R$ pair
production. An exception occurs at low $m_0$ values where selectron
pair production is dominant, but where $m_{\te_R}\simeq m_{\tz_1}$. 
Then, the mass gap
$m_{\te_R}-m_{\tz_1}$ becomes so small that there was very little
visible energy in the slepton pair events, resulting in
very low detection efficiency, causing a turnover in the reach contours.
For $\tan\beta$ values larger than those explored in Ref. \cite{bmt}, 
the large $\tau$ Yukawa coupling makes
$\ttau_1$ significantly lighter than $\te_R$  so that
close to the boundary of excluded region at small $m_0$ 
$\te_R\bar{\te}_R$ events may still be observable, 
while $\ttau_1\bar{\ttau_1}$ events are not.
\subsection{Updated reach results}

In this section, we update our earlier
reach projections\cite{bmt} for a linear collider.
We present reach projections for linear colliders with 
$\sqrt{s}=0.5$ and 1~TeV of energy in the CM frame. 
We also expand the
range of $m_{1/2}$ (to 1.6 TeV) and $m_0$ (to 8 TeV) beyond the values
presented in Ref. \cite{bmt}. This allows us to explore the
entire stau co-annihilation strip, 
the $A$-annihilation funnel and the HB/FP region. 
In our analysis, we restrict ourselves to the trilinear SSB term $A_0=0$.
For the most part, our results are qualitatively insensitive to variations 
in $A_0$. An exception occurs for particular $A_0$ choices which may greatly
reduce the value of $m_{\tst_1}$, and lead to a top squark-neutralino
co-annihilation region\cite{stop}.

Our first results are presented in Fig.~\ref{fig:30plc}, 
where we show the $m_0\ vs.\ m_{1/2}$ plane for $\tan\beta =30$, 
$\mu >0$ and $A_0=0$. The left-most red region at low $m_0$ is disallowed 
because the LSP would become a stau.
The right-most red region (large $m_0$) is mainly excluded by a 
lack of appropriate REWSB, although this includes as well points
with no convergent RGE solution as generated by ISAJET.
The precise location of the boundary of the large $m_0$ red region 
depends somewhat on the computer code, and also on the assumed fermion 
masses.
The lower yellow region is excluded by LEP2 chargino 
searches, which require $m_{\tw_1}>103.5$~GeV. In addition, the region 
below the yellow contour gives a light (SM-like) Higgs boson with
$m_h<114.4$~GeV. 

Using the standard dilepton cuts as described above, we find a SM
background level of $\sigma_{SM}=1.79$ fb (0.045 fb) for a LC with
$\sqrt{s}=0.5$ TeV (1.0 TeV) with right polarized beams using $P_L(e^-
)=-0.9$.\footnote{In our assessment of SM backgrounds, we have evaluated
only the backgrounds from $2\to 2$ processes.} It has been shown in
Ref. \cite{jlc1} that backgrounds from $2 \to 3$ processes such as
$e^+e^- \to \nu\nu Z, e^+e^- Z$ or $e^{\pm}\nu W^{\mp}$ production or 
$2 \to 4$ processes such as $e^+e^- \to W^+W^-$ production are also
efficiently removed by these cuts, at least for
$\sqrt{s}=500$~GeV. Although the cross sections for these processes grow
with energy, we expect that these cuts will remove the {\it bulk} of
these backgrounds also at $\sqrt{s}=1000$~GeV; for instance, the
dominant portion of the $eeWW$ background that comes from
``$\gamma\gamma$'' collisions will be removed by the acoplanarity and
other cuts; much of the remaining cross section will have $p_T(WW) \leq
M_W$ and will also be reduced, though not eliminated by these cuts.
However, since these backgrounds have not been included in our
evaluation it is possible that the statistical significance of the
signal may be somewhat over-estimated for $\sqrt{s}=1000$~GeV.  The
regions to the left of the lower (upper) blue contour yield a
supersymmetric signal at the $5\sigma$ level assuming 100 fb$^{-1}$ of
integrated luminosity at the 0.5 (1) TeV LC. An increased reach for
slepton pairs may be obtained by searching for ditau events originating
from stau pair production.  While we did not perform a detailed
signal-to-background analysis for this channel, we do show the
kinematic limit for stau pair production by a dashed light-blue contour,
which for $\tan\beta =30$, lies somewhat above the blue dilepton reach
contour.
Thus, a significant gain in reach might be acquired by
searching for acollinear di-tau events arising from stau pair production.
In particular, this signal may give access to the large $m_{1/2}$
part of the stau co-annihilation region.

The green contour denotes the reach of a 0.5 or 1~TeV LC for SUSY via
the $1\ell +2j+\eslt$ channel arising from chargino pair production,
using the standard cuts given above. Unlike Ref.\cite{bmt}, we use
$P_L=0$, and find a background level of 15.5 fb (2.1 fb) for
$\sqrt{s}=0.5$ (1)~TeV (the beam polarization is not important for this
reach contour).  For most of parameter space, the reach contours follow
closely along the $m_{\tw_1}=250$ (500)~GeV mass contours, indicating
that chargino pair production can be seen with standard cuts almost to
the kinematical limit for chargino pair production.
\FIGURE{\epsfig{file=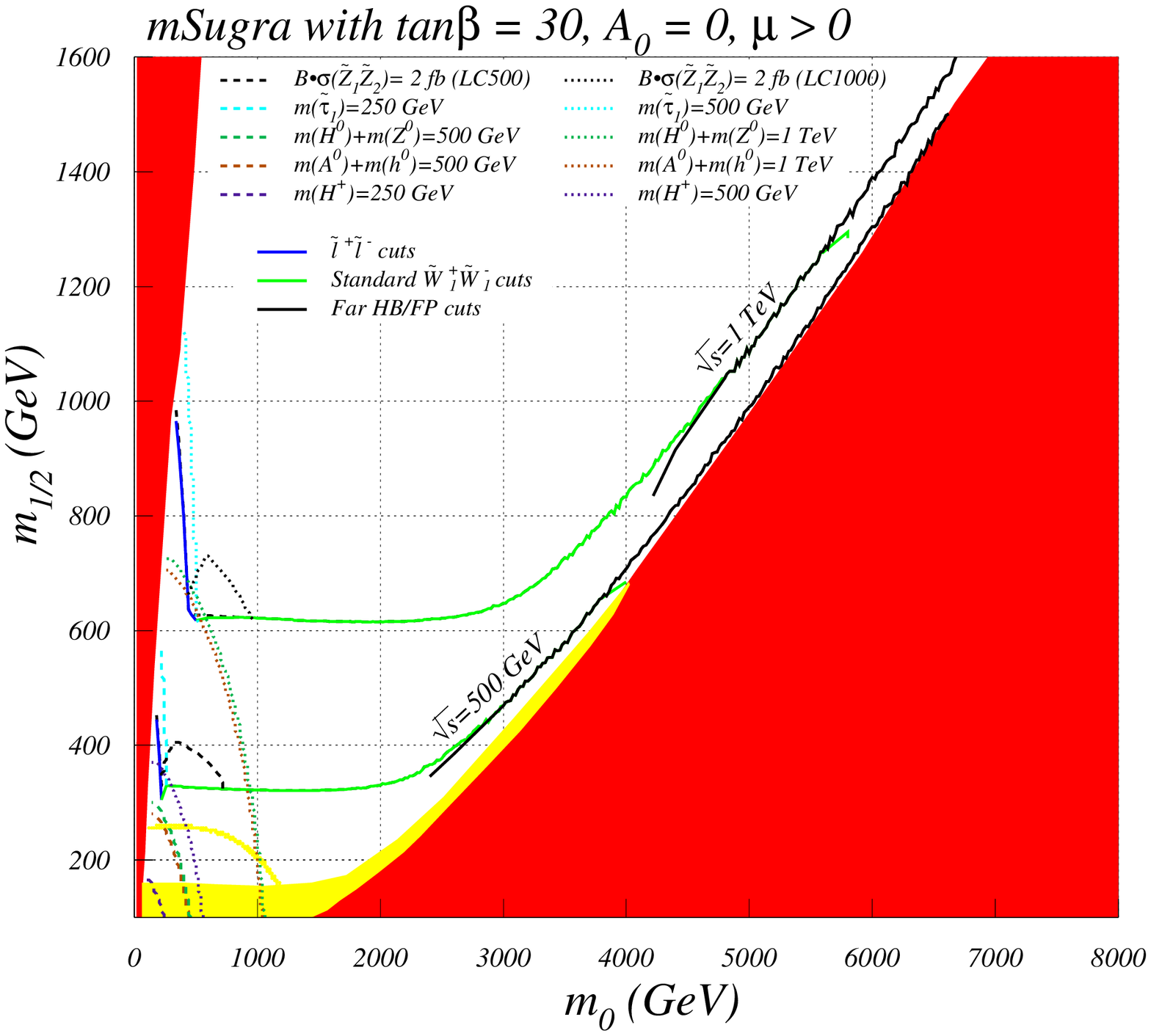,width=14cm} 
\vspace*{-0.8cm}
\caption{Reach of a linear collider for supersymmetry in the mSUGRA
model for $\sqrt{s}=500$ and $1000$~GeV, 
for $\tan\beta =30$, $A_0=0$ and $\mu >0$. The reach via slepton pair 
production is denoted by the blue contour, while standard cuts for 
chargino pair production yield the green contour. Special chargino pair
cuts yield the black contour in the HB/FP region. The red region is 
theoretically excluded, while the yellow region 
is excluded by LEP2 measurements. Below the yellow contour, $m_h \leq
114.4$~GeV. 
}
\label{fig:30plc}
}
An exception occurs when $m_0$ becomes
very large, in the HB/FP region. Around $m_0\sim 4000$~GeV 
($m_0\sim 6000$~GeV) for $\sqrt{s}=0.5$ TeV (1~TeV), the reach contour departs
from the kinematic limit. The termination of the reach contour
occurs because in this region, the superpotential parameter $\mu$
becomes very small, and the light chargino $\tw_1$ and neutralino $\tz_1$
become higgsino-like, and increasingly mass degenerate. The $Q$-value from
$\tw_1\to \tz_1 f\bar{f}'$ decay (the $f$s are light SM fermions)
becomes very small, and very little visible energy is released by the 
chargino decays. This causes the detection efficiency for 
$1\ell +2j +\eslt$ events to decrease sharply, 
leading to a corresponding reduction in the reach
using the standard cuts. 

To understand what is happening in the HB/FP region, we
show relevant sparticle  masses in Fig.~\ref{fig:30p225}{\it a})
for $m_{1/2}=225$~GeV, $\tan\beta =30$, $A_0=0$ and $\mu >0$ versus
the parameter $m_0$.
As $m_0$ varies from 1400~GeV to nearly 2200 GeV, 
{\it i.e.} as we approach the HB/FP region for fixed $m_{1/2}$
with increasing $m_0$. As $m_0$ increases, $|\mu |$ is seen to be decreasing.
Since the value of $SU(2)$ gaugino mass $M_2$ is essentially fixed, 
the various chargino and neutralino masses also decrease, with the lighter
ones becoming increasingly higgsino-like. The plot is terminated when
the LEP2 limit $m_{\tw_1} \ge 103.5$~GeV 
is reached. Of great importance is that
the $\tw_1 -\tz_1$ mass gap is also decreasing, although in this case
it remains substantial out to the edge of parameter space.

In Fig.~\ref{fig:30p225}{\it b}), we show the total cross section 
for various chargino and neutralino production reactions versus $m_0$ 
as in frame {\it a}). At the lower $m_0$ values, $\sigma (\tw_1^+\tw_1^-)$
pair production dominates the total SUSY production cross section.
As $m_0$ increases, and $|\mu |$ decreases, the other charginos and 
neutralinos become light as well, and many more reactions ``turn on''
in the HB/FP region. Although we will focus mainly on $\tw_1^+\tw_1^-$
pair production, it is important to note that
many SUSY production reactions can occur in the HB/FP region,
and can lead to an assortment of SUSY events from the
production and cascade decays of the heavier chargino and neutralino
states.
\FIGURE{\epsfig{file=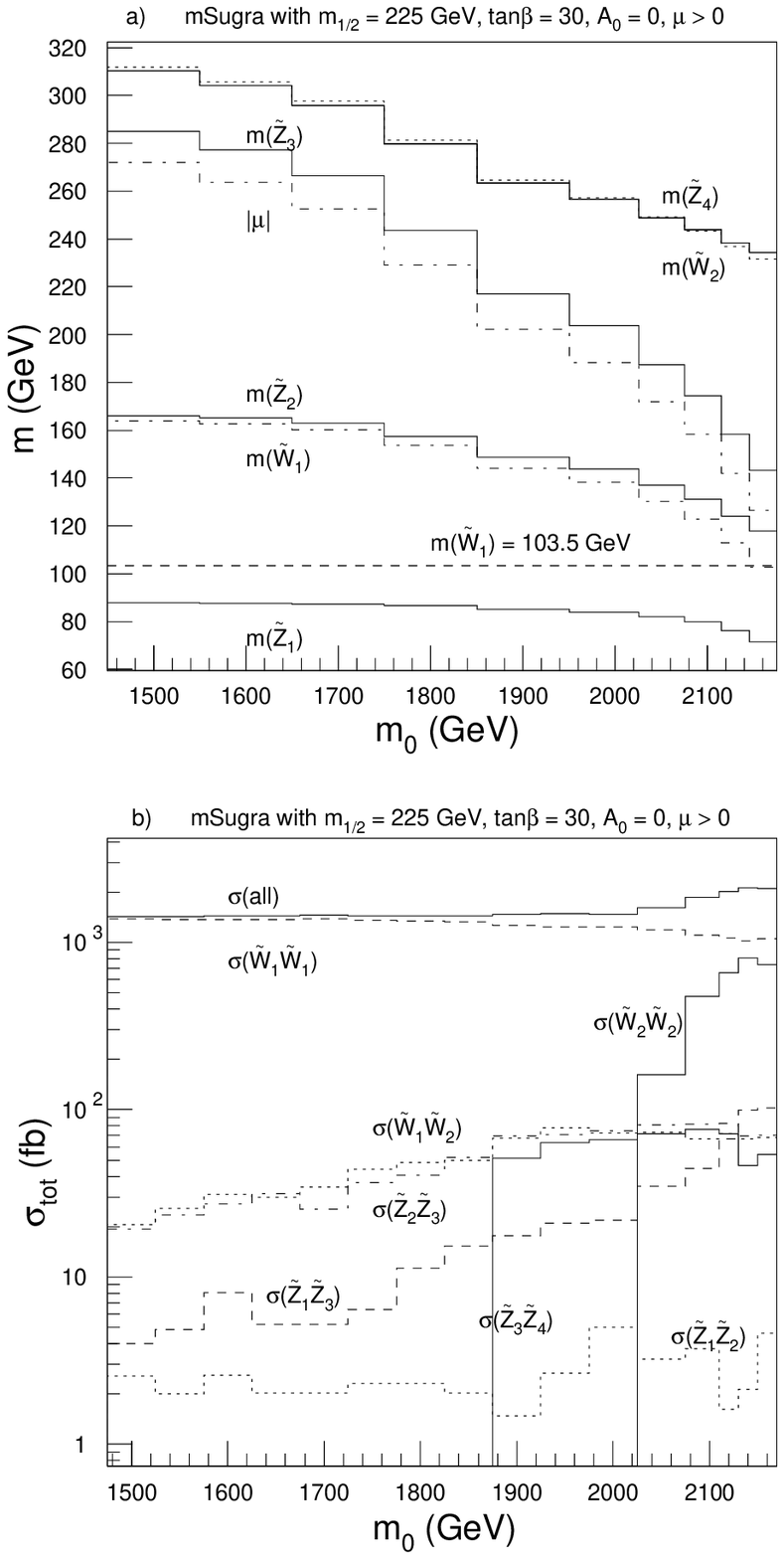,width=14cm}
\vspace*{-0.8cm}
\caption{Plot of {\it a}). sparticle masses and {\it b}).
sparticle pair production cross sections 
versus $m_0$ in the
HB/FP region for $m_{1/2}=225$~GeV, $\tan\beta =30$, $A_0=0$ and
$\mu >0$ for a $\sqrt{s}=500$~GeV $e^+e^-$ collider.
}
\label{fig:30p225}
}

A similar plot is shown in Fig.~\ref{fig:30p900}, except this time
for $m_{1/2}=900$~GeV, {\it i.e.} in the upper regions of the
hyperbolic branch. In this case, $M_2$ is much larger than the case 
\FIGURE{\epsfig{file=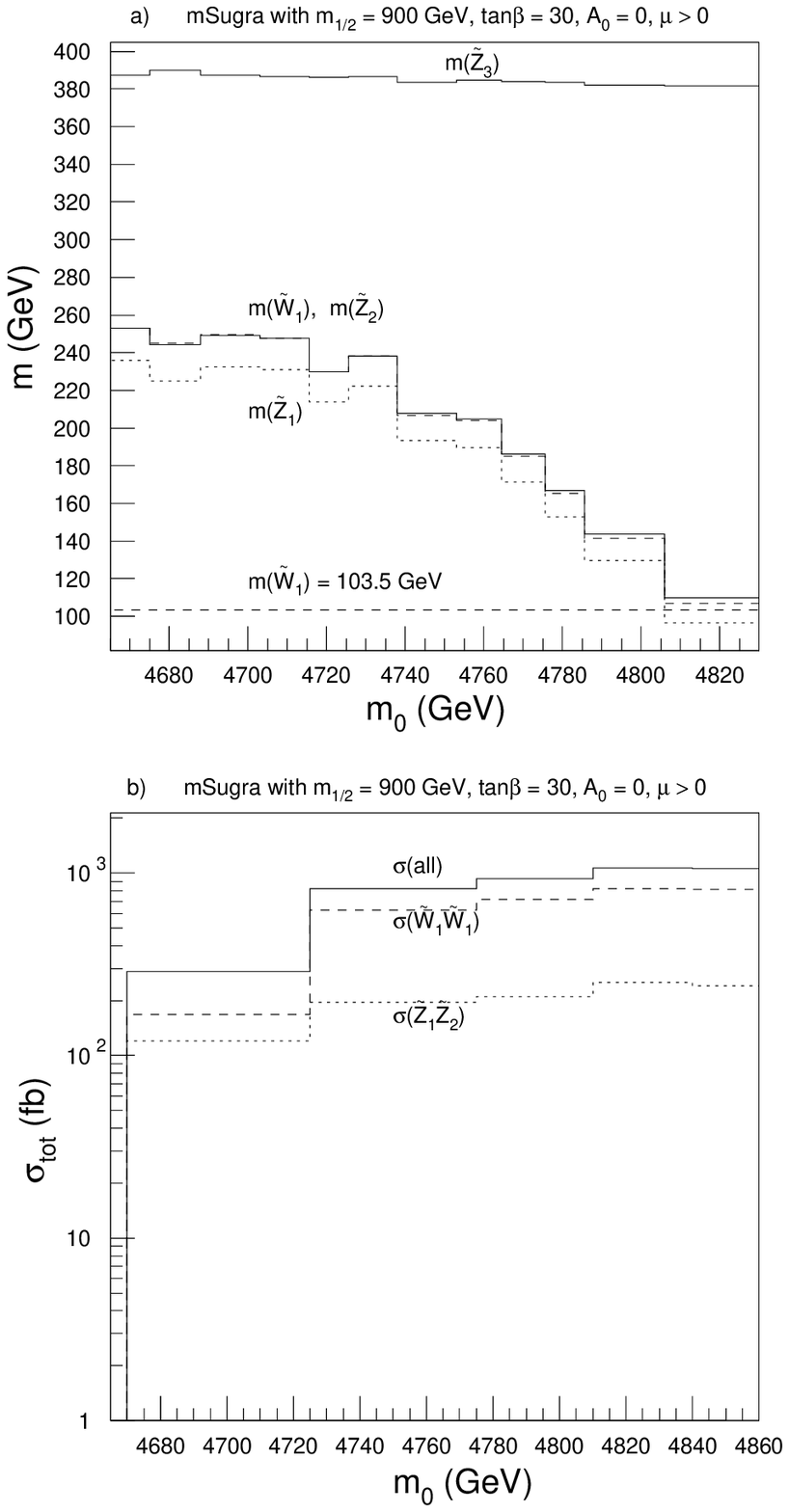,width=14cm} 
\vspace*{-0.5cm}
\caption{Plot of {\it a}). sparticle masses and {\it b}).
sparticle pair production cross sections versus $m_0$ in the
HB/FP region for $m_{1/2}=900$~GeV, $\tan\beta =30$, $A_0=0$ and
$\mu >0$ for a $\sqrt{s}=500$~GeV $e^+e^-$ collider.
}
\label{fig:30p900}}
shown in Fig.~\ref{fig:30p225}, and so the heavier charginos and neutralinos
remain inaccessible to a LC. As $m_0$ increases, again
$m_{\tw_1}$ and $m_{\tz_1}$ decrease.\footnote{Since $|\mu|$ decreases
very rapidly as $m_0$ increases and approaches the theoretical
boundary of the HB/FP region, its precise value is not easy to obtain using
numerical methods. The value of $\mu$, of course, directly affects the 
chargino and neutralino masses. In this figure, we have smoothed out
what appeared to be rather large numerical fluctuations in two of the bins.} 
 But in this case the $\tw_1 -\tz_1$
mass gap is much smaller, reaching only several~GeV at the limit of
parameter space. Clearly, in this upper $m_{1/2}$ region of the 
hyperbolic branch, there will be little visible energy
emerging from chargino 3-body decays, making detection of chargino pair
events difficult using standard cuts. In addition, as shown in frame 
{\it b}), only $\tw_1^+\tw_1^-$ and $\tz_1\tz_2$ pair 
production occur, so fewer anomalous events are expected in the upper
HB/FP region. Since $m_{\tz_2}\sim m_{\tw_1}$,
there will also be little visible energy from $\tz_2\to\tz_1 f\bar{f}$
decay, so that $\tz_1\tz_2$ production will also be more difficult to observe.
In the deep HB/FP region where $|\mu| \ll M_{1,2}$, one of the neutralinos
is mainly higgsino-like with roughly equal components of ${\tilde h}_u$ and
${\tilde h}_d$. The other neutralinos, being orthogonal to these, thus
either have equal magnitudes for their ${\tilde h}_u$ and ${\tilde h}_d$
content, or this content is small. In either case, the $Z\tz_i\tz_i$
coupling is dynamically suppressed\cite{dkt} in this region. This accounts
for the strong suppression of $\sigma ({\tz_2\tz_2})$ (recall that the
electron sneutrino is very heavy) in Fig.~\ref{fig:30p900}{\it b}.

Coincidentally,
the reach of a $\sqrt{s}=0.5$~TeV LC in the
$1\ell +2j +\eslt$ channel 
using the standard cuts terminates in the HB/FP region
at nearly the same $m_{1/2}$ value as does the reach of the CERN LHC
shown in Ref. \cite{bbbkt}, and again in Sec.~4 of this paper. 
Meanwhile, the contour of chargino pair kinematic accessibility extends to
much higher $m_{1/2}$ values, along the hyperbolic branch.
This motivated us to examine strategies to extend 
the reach of a LC
to the large $m_{1/2}$ part of the HB/FP region.

To find better suited
signal selection cuts for the HB/FP region, we examine
a particular case study for the mSUGRA point which is beyond the
projected reach of the LHC\cite{bbbkt}: 
\be
m_0,\ m_{1/2},\ A_0,\ \tan\beta ,\ sign (\mu ) =
\ 4625\ {\rm~GeV},\ 885\ {\rm~GeV}, 0,\ 30,\ +1 ,
\ee
for which various sparticle masses and 
parameters are listed in Table~\ref{tab:cs1}. We will refer to this as
case 1. Not only is this point inaccessible at the LHC, but 
most of the sparticles are also inaccessible to a LC,
with the exception being the lighter charginos and neutralinos.
While $m_{\tw_1}=195.8$~GeV, so that $\tw_1^+\tw_1^-$ pair production
occurs at a large rate at a $\sqrt{s}=0.5$ TeV $e^+e^-$ collider, 
the $\tw_1 -\tz_1$ mass gap is only 14.2~GeV, so little visible energy
is released in chargino pair production events.

With this in mind, we generate SUSY events for this case study
using ISAJET 7.69 for a linear collider with $\sqrt{s}=0.5$ TeV and
unpolarized beams, 
including bremsstrahlung and beamstrahlung
for background events.  
The beamstrahlung parameters,
defined in Ref. \cite{PChen},
are taken to be $\Upsilon =0.1072$ with beam length $\sigma_z =0.12$~mm.


\TABLE{
\begin{tabular}{lc}
\hline
parameter & value (GeV) \\
\hline
$M_2$ & 705.8 \\
$M_1$ & 372.2 \\
$\mu$ & 185.9 \\
$m_{\tg}$ & 2182.7 \\
$m_{\tu_L}$ & 4893.9 \\
$m_{\te_L}$ & 4656.1 \\
$m_{\tw_1}$ & 195.8 \\
$m_{\tw_2}$ & 743.5 \\
$m_{\tz_1}$ & 181.6 \\
$m_{\tz_2}$ & 196.2 \\ 
$m_{\tz_3}$ & 377.3 \\
$m_{\tz_4}$ & 760.0 \\
$m_A$ & 3998.3 \\
$m_h$ & 122.0 \\
$\Omega_{\tz_1}h^2$& 0.0104\\
$BF(b\to s\gamma)$ & $3.34\times 10^{-4}  $\\
$\Delta a_\mu    $ & $0.6 \times  10^{-10}$\\
\hline
\label{tab:cs1}
\end{tabular}
\caption{Masses and parameters in~GeV units for case 1 
for $m_0,\ m_{1/2},\ A_0,\ \tan\beta ,\ sign\mu =$
4625~GeV, 885~GeV, 0, 30, +1 in the mSUGRA model.
The spectrum is obtained using ISAJET v7.69.}
}

In Fig.~\ref{fig:dist}{\it a}) we show the distribution of
$E_{visible}$ from events with 1-lepton and two jets expected at a
$\sqrt{s}=500$~GeV LC. 
The solid black histogram represents the SUSY case study, 
which peaks at very low $E_{visible}$, as expected from the low energy release
from chargino decays. The small number of events around $E_{visible} =
250$~GeV is from $Zh$ production. 
\FIGURE{\epsfig{file=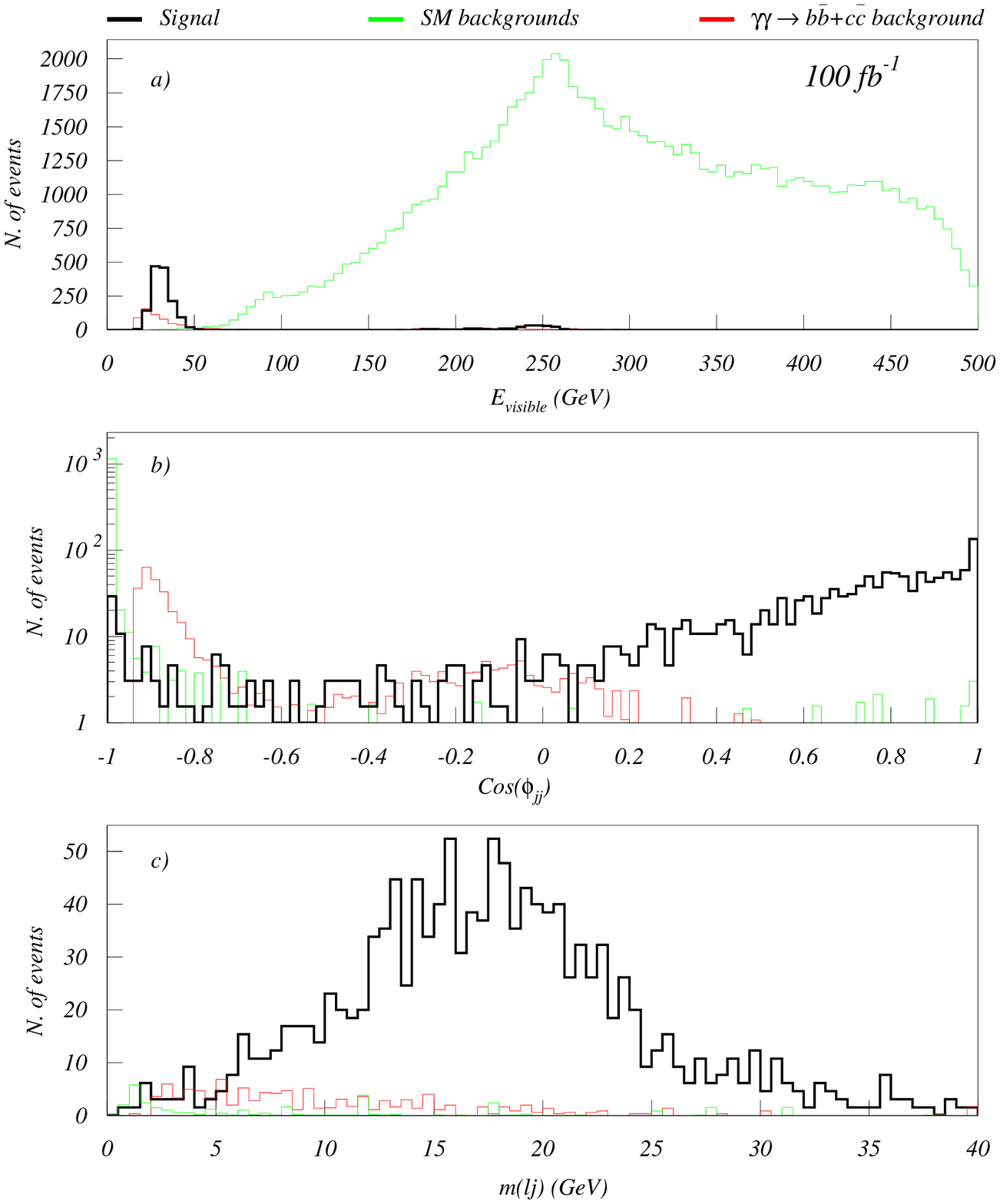,width=14cm} 
\vspace*{-0.8cm} 
\caption{Distribution in {\it a}) $E_{visible}$ for mSUGRA signal
(black histogram) with 
($m_0,\ m_{1/2},\ A_0,\ \tan\beta,\ sign(\mu )=4625$~GeV, 885~GeV, 0, 30, 1)
after cuts  in the first row of Table~\ref{tab:cuts}. 
We take $\sqrt{s}= 500$~GeV and an integrated
luminosity of 100 fb$^{-1}$,
and adopt beamstrahlung parameters $\Upsilon =0.1072$ and 
$\sigma_z=0.12$ mm.
The ISAJET SM background is shown by the green histogram, while
the background from $\gamma\gamma\to c\bar{c},\ b\bar{b}$ is shown in red.
In {\it b}), we show the distribution in transverse plane dijet opening angle
requiring, in addition, that 
20~GeV$<E_{visible}<100$~GeV. In {\it c}), we show the
distribution in $m(\ell j)$, after the aditional requirement 
$\cos\phi (jj)>-0.6$.
The jet entering the $m(\ell j)$ distribution is the one that is 
closest in angle to the
lepton direction.
\label{fig:dist}
}
}
The green histogram shows the sum of all $2\to2$ SM backgrounds
as generated by ISAJET. The large $E_{visible}$ component of these
arises from $WW$, $ZZ$ and $t\bar{t}$ production, where some
energy is lost due to associated neutrino emissions. The SM background 
distribution extends to low $E_{visible}$ values, and has a visible
shoulder at $E_{visible}\sim M_Z$ due to processes such as 
$e^+e^-\to Z\to b\bar{b}, c\bar{c}$, where the $Z$ can be made by 
convoluting the subprocess reaction with the electron PDF, and the
lepton from the decay of the heavy flavor is accidently isolated.
The bulk of the $2\to 2$ 
SM background can be eliminated by requiring low values of
$E_{visible}$. In this case we require 
\be
20\ GeV <E_{visible}< 100\ GeV.
\ee
The upper limit is chosen to be well above the case 1 signal distribution 
endpoint to accommodate later scans over all mSUGRA parameter space, 
including points which allow larger $\tw_1 -\tz_1$ mass gaps, and
somewhat harder $E_{visible}$ distributions. We also show a red 
histogram which shows the results of the evaluation of the background 
from
$e^+e^-\to e^+e^-  c\bar{c},\ e^+e^-  b\bar{b}$ processes
when both initial leptons escape detection
when being scattered at a very small angle.
We evaluated this background, which mainly arises from photon photon
collisions,  
using the PYTHIA event generator\cite{pythia}.
In the case of $b\bar{b}$ production, the isolated
lepton arises from semi-leptonic $b\to c\ell\nu$ decay, and the jets come
one from a $b$ quark, and the other from the charm quark.

In $\gamma\gamma\to b\bar{b}$ events, the $b$ and $\bar{b}$
will typically emerge back-to-back in the transverse plane.
Thus, in Fig.~\ref{fig:dist}{\it b}) we plot the distribution in
$\cos\phi (jj)$, where $\phi (jj)$ is the transverse dijet opening angle.
The signal is distributed over a range of $\cos\phi (jj)$ values, and actually
peaks at $\cos\phi (jj)\sim 1$. The background peaks at $\cos\phi (jj)\sim -1$,
so we require a cut of 
\be
\cos\phi (jj)> -0.6 .
\ee

Finally, any surviving background arising from $b\bar{b}$ or
$c\bar{c}$ production followed by semileptonic heavy flavor decay 
is likely to 
have a jet-lepton invariant mass bounded by the heavy flavor
mass (at least at parton level).
In Fig.~\ref{fig:dist}{\it c}) we show the distribution in 
$m(\ell j)$ where we form the invariant mass from the jet which is closest 
to the isolated lepton in space angle. 
Some additional background removal at low cost to signal is gained by
requiring
\be
m(\ell j_{near}) >5\ GeV.
\ee

At this point, the distribution is clearly dominated by
signal.
The cross sections in fb after each cut for
signal and background are shown in Table~\ref{tab:cuts}, where we
include in addition background from the $2\to 4$ process $e^+e^-\to
\ell\nu q\bar{q}'$ (evaluated using CompHEP\cite{comphep}). In the $2\to 4$
calculation we eliminate Feynman diagrams such as $WW$ pair production
which are already accounted for as $2\to 2$ processes in ISAJET.  The
$2\to 4$ processes are negligible after the $E_{visible}$ cut.  For all
frames of Fig.~\ref{fig:dist} we have assumed 100 fb$^{-1}$ total
integrated luminosity. We recognize that the dominant background to the 
rather soft signal which comes from two photon collisions will be sensitive
to the beamstrahlung parameters, and hence to the shape of the
beams. Nevertheless, the large signal to background ratio that we find
encourages us to expect that our conclusions about the large reach of LC
will be qualitatively unaltered.

\TABLE{
\begin{tabular}{lcccc}
\hline
cuts & case 1 & ISAJET BG & $\gamma\gamma\to c\bar{c},b\bar{b}$ & 
$\ell\nu q\bar{q}'$  \\
\hline
$\eta,\ E,\ \Delta R$ & 16.2 & 897.1 (483) & 9.2 (6.2) & 448 (712)  \\
$20~GeV<E_{vis}<100$~GeV & 14.4 & 12.6 (3.5) & 5.4 (4.9) & 0.16 (0.08)  \\
$\cos\phi (jj)>-0.6$ & 13.5 & 0.34 (0.2) & 1.1 (1.1) & 0.04 (0.02) \\
$m(\ell j)>5$~GeV & 12.9 & 0.17 (0.1) & 0.8 (0.8) & 0.04 (0.02) \\ 
\hline
\label{tab:cuts}
\end{tabular}
\vspace*{-0.4cm}
\caption{Cross section after cuts in fb for mSUGRA case 1 signal and
ISAJET SM backgrounds, two photon background $\gamma\gamma\to c\bar{c},\ 
b\bar{b}$
and the $2\to 4$ process $e^+e^-\to \ell\nu_\ell q\bar{q}'$.
We take $\sqrt{s} =0.5$ TeV collider CM energy. The corresponding 
background for $\sqrt{s}= 1$~TeV case is listed in 
parenthesis.}
}
We now require a $5\sigma$ signal for SUSY events above the total
SM background as listed in Table~\ref{tab:cuts}, for 100 fb$^{-1}$
of integrated luminosity and  scan mSUGRA points in the HB/FP region of
Fig.~\ref{fig:30plc}. The new result is 
the black contour, below which the mSUGRA parameter space is accessible 
by the LC at 5$\sigma$ level.
One can see that this contour pushes the reach of the LC to much higher values
of $m_{1/2}$.
The contour 
peters out at low $m_{1/2}$ values, where the visible energy arising from
chargino pair production is typically much higher than the 100~GeV maximum
required by our cuts. However, this low $m_{1/2}$ region is already well
covered by the standard chargino search cuts listed at the beginning 
of this section. 
The new cuts for the far HB/FP region work for 
$e^+e^-$ colliders with $\sqrt{s}=1$ TeV as well.
The $\sqrt{s}=1$ TeV reach contour extends even beyond the
limits of parameter space shown in Fig.~\ref{fig:30plc}.

To complete our SUSY reach contours, we also examined the 
reach of a LC for SUSY via the $e^+e^-\to \tz_1\tz_2$ reaction
in the non-HB/FP part of parameter space, where neither
chargino pair production nor slepton pair production is
kinematically accessible. In this case, we follow Ref. \cite{bmt}
in asking for $b\bar{b}+\eslt$ events from $e^+e^-\to \tz_1\tz_2$
production followed by $\tz_2\to \tz_1 h$, where $h\to b\bar{b}$.
In Ref. \cite{bmt}, no background was found after a series of cuts
listed at the beginning of this section. Here, we do not perform
complete event generation at every point in parameter space, but instead
require that 
\be
\sigma (e^+e^-\to\tz_1\tz_2 )\times BF(\tz_2\to\tz_1 h)> 2\ {\rm fb} ,
\ee
which should yield $\sim 10$ signal events for 100 fb$^{-1}$ of integrated 
luminosity assuming an efficiency of $5-6\%$ as found in Ref. \cite{bmt}.
The resulting reach contour is shown in Fig.~\ref{fig:30plc} as the black
contour linking the slepton pair reach to the chargino reach contour.
It gives some additional parameter space reach to a LC, although it is
in a dark matter {\it disfavored} region of parameter space (unless
$\tan\beta$ is large, and the $H,A$-annihilation funnel cuts through
it). 
There is a 
turnover in the $e^+e^-\to\tz_1\tz_2$ reach contour at low $m_0$;
this occurs because in this region, $\tz_2\to\ttau_1\bar{\tau}$ decay
becomes accessible, resulting in a suppression of the $\tz_2\to\tz_1 h$
branching fraction.
However, the $\tz_2\to\ttau_1\bar{\tau}$ signal may also be detectable; if so,
the gap caused by the turn-over just mentioned would be filled.

Finally, we show the kinematic limit for $e^+e^-\to ZH$ (green-dashed or 
dotted contours),
$e^+e^-\to Ah$ (orange dashed or dotted contours) and $e^+e^-\to H^+H^-$
(purple dashed or dotted contours). The dashes are for $\sqrt{s}=0.5$
TeV, while dotted are for $\sqrt{s}=1$ TeV. For $\tan\beta =30$,
these contours always lie below the sparticle reach contours,
so if heavier SUSY Higgs bosons are seen (at least in the channels 
mentioned above), 
sparticles should also be seen if SUSY is realized as in the mSUGRA 
framework.
 
In Fig.~\ref{fig:10plc}, we show LC reach contours for the same 
mSUGRA parameter plane as in Fig.~\ref{fig:30plc}, except this time
$\tan\beta =10$. Qualitatively, many of the reach contours are similar to the
$\tan\beta =30$ case of Fig.~\ref{fig:30plc}, and in particular, the new cuts
\FIGURE{\epsfig{file=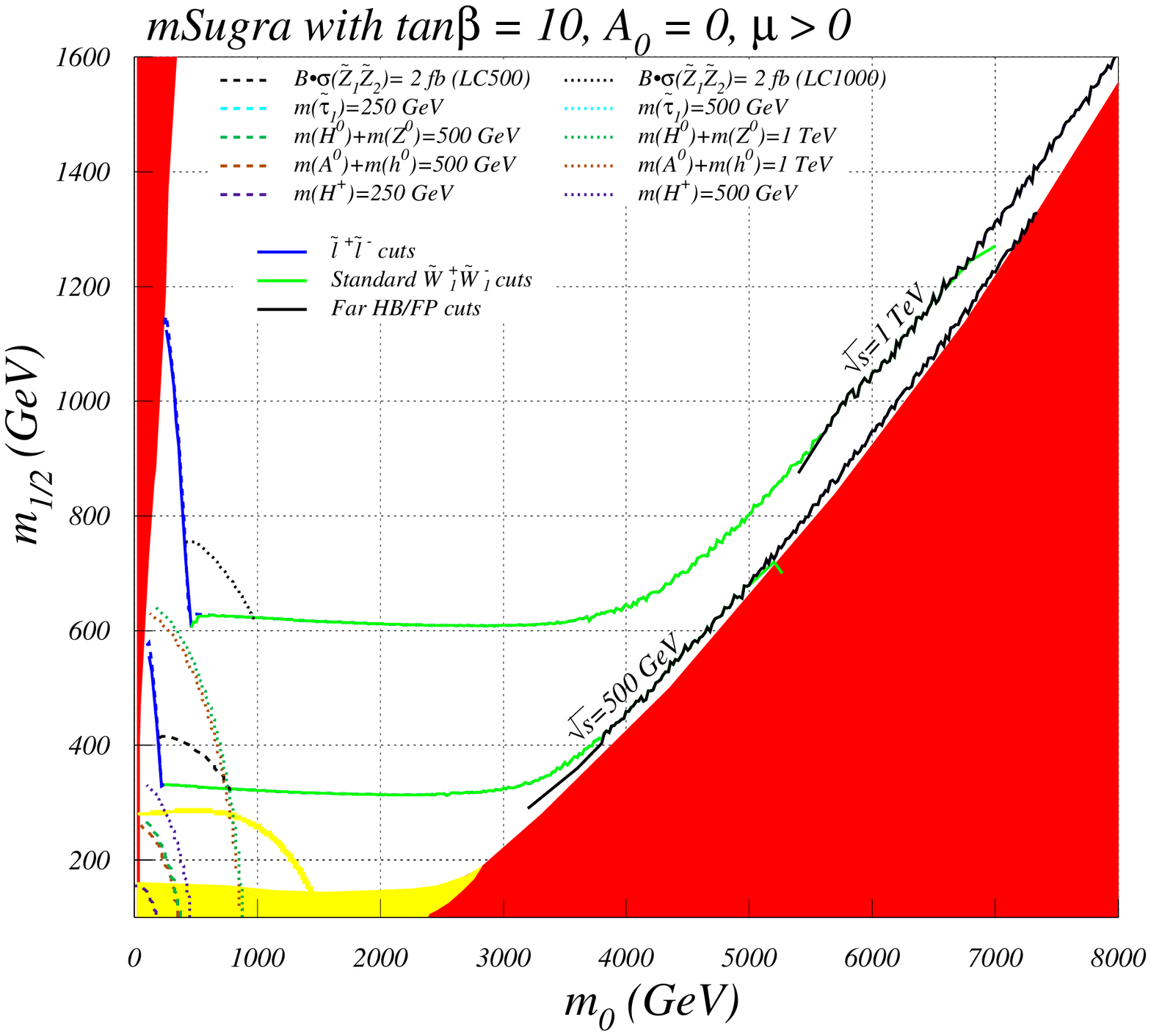,width=14cm} 
\vspace*{-0.5cm} 
\caption{Reach of a linear collider for supersymmetry in the mSUGRA
model for $\sqrt{s}=500$ and $1000$~GeV, 
for $\tan\beta =10$, $A_0=0$ and $\mu >0$. The colors on the various
regions and on the different contours are as in Fig.~\ref{fig:30plc}.
}
\label{fig:10plc}}
designed to access the far HB/FP region again allow the LC reach to extend
into the high $m_{1/2}$ section of the hyperbolic branch.
One difference for $\tan\beta =10$ results is that the $\ttau_1^+\ttau_1^-$
kinematic reach contour now lies nearly atop the selectron/smuon
reach contour using the dilepton cuts described at the beginning of 
this section. 

In Fig.~\ref{fig:45mlc}, we show the same reach contours in the 
$m_0\ vs.\ m_{1/2}$ plane, but this time for $\tan\beta =45$ and $\mu <0$.
The standard slepton pair and chargino pair production reach contours 
are similar to the low $\tan\beta$ cases. In this case, the 
far HB/FP region cuts allow the reach to be extended, although
the reach contour terminates in the part of the red region
where the numerical solutions
to the renormalization group equations do not converge as per
\FIGURE{\epsfig{file=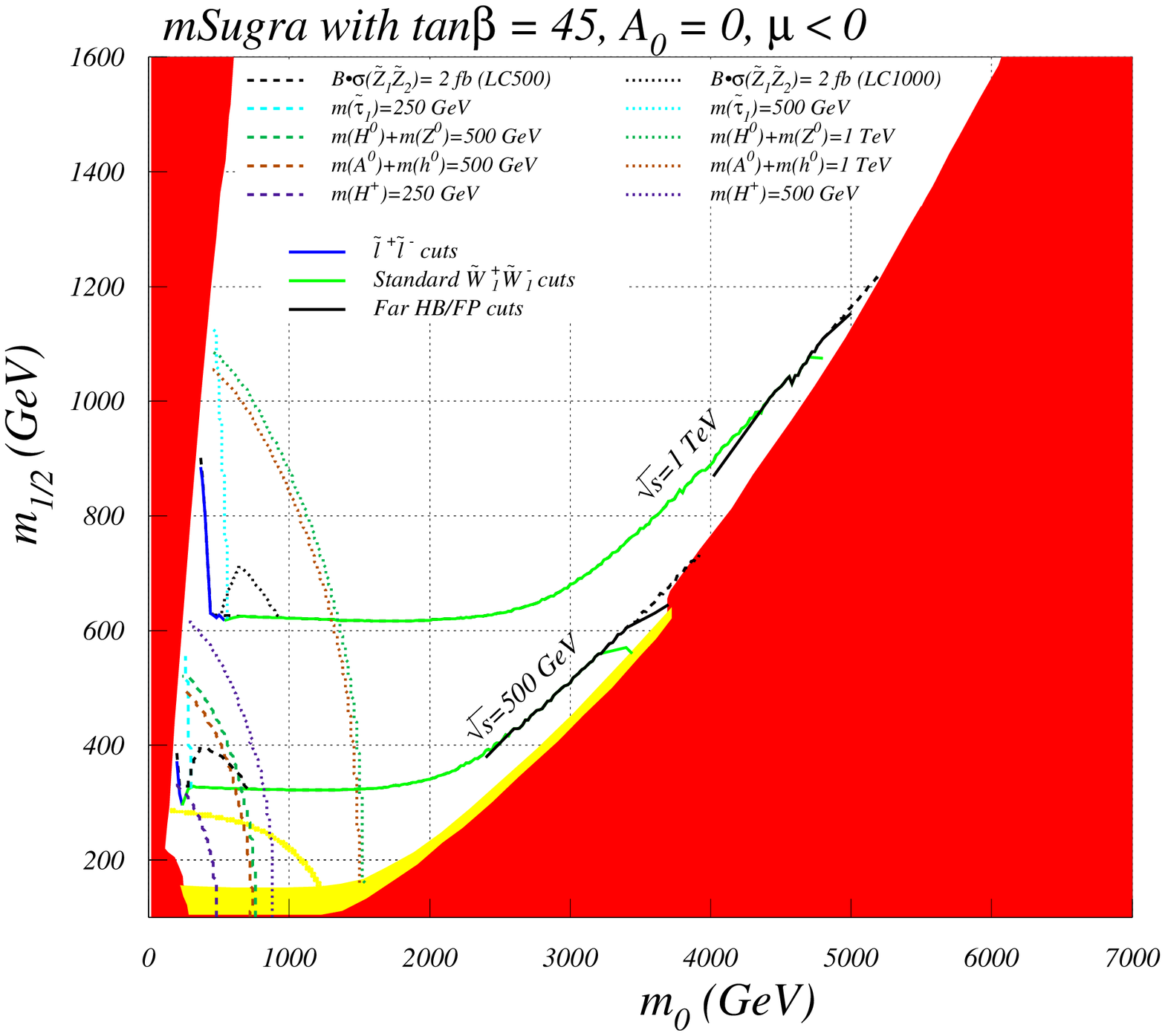,width=14cm} 
\vspace*{-1cm} 
\caption{Reach of a linear collider for supersymmetry in the mSUGRA
model for $\sqrt{s}=500$ and $1000$~GeV, 
for $\tan\beta =45$, $A_0=0$ and $\mu <0$. The colors on the various
regions and on the different contours are as in Fig.~\ref{fig:10plc}.
}
\label{fig:45mlc}}
the criteria in ISAJET.
A tiny region remains inaccessible to our new cuts, between the black 
solid contours  and the dashed contours that depicts the kinematic limit
for chargino pair production.
Another feature of the $\tan\beta =45$ plot is that the stau pair
kinematic region has expanded even more beyond where signals for $\te_R$
and $\tmu_R$ may be accessible in the dilepton channel.
In addition, for this large value of $\tan\beta$, the
heavy Higgs bosons are much lighter than the low $\tan\beta$
cases\cite{ltanb}, 
and now there exist regions of parameter space where $ZH$ and
$Ah$ production may be accessible, while sparticles are not: 
indeed such a situation would point to a large value of $\tan\beta$
which would of course be independently measureable from the properties of
the detected Higgs bosons\cite{barger}. 
These regions
all occur well below the reach of the LHC for SUSY, which will be shown 
in Sec.~4. Thus, if nature chooses $\tan\beta =45$ in an mSUGRA-like
model, and the LC sees only Higgs bosons beyond the SM, it is likely that
the existence of SUSY would already have been
established by LHC experiments. 

In Fig.~\ref{fig:52plc}, we show our final LC reach plot, taking
$\tan\beta =52$ with $\mu >0$. In this case, there is still substantial
reach for chargino pairs via the standard cuts, and the special cuts for
the far HB/FP region again allow extended reach into this area. For this
large a value of $\tan\beta$, the dilepton reach contour has been
completely consumed by the expanding forbidden region on the left where
$\ttau_1$ becomes the LSP.  In addition, the $e^+e^-\to\tz_1\tz_2$ reach
region has shrunk due to the increased branching fraction for
$\tz_2\to\ttau_1\tau$ decay.  However, the kinematic reach for stau
pairs has greatly increased, and becomes especially important for very
large $\tan\beta$, especially if nature has chosen to reduce LSP
dark matter
from the early universe via co-annihilation with staus.
\FIGURE{\epsfig{file=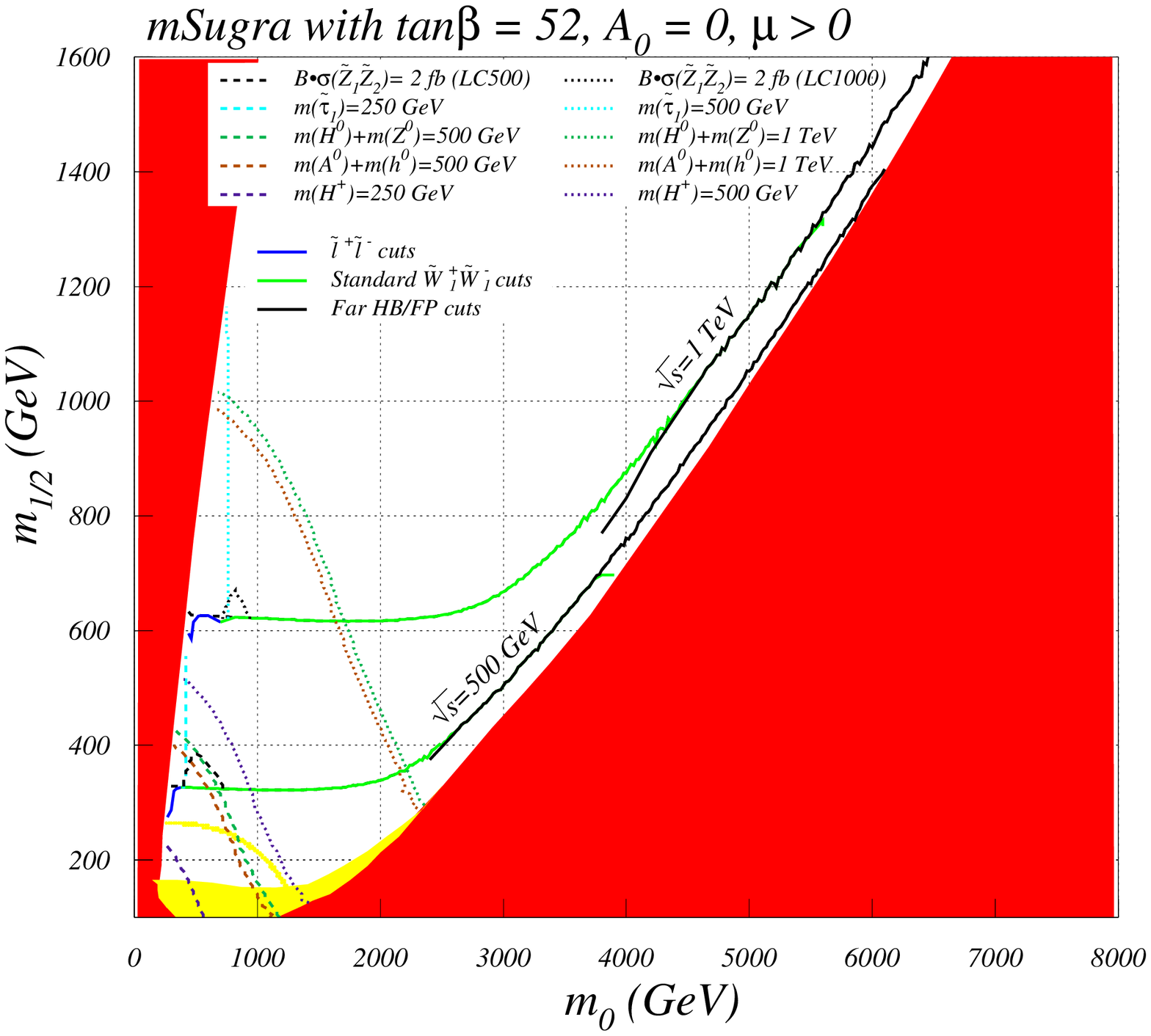,width=14cm} 
\vspace*{-1cm} 
\caption{Reach of a linear collider for supersymmetry in the mSUGRA
model for $\sqrt{s}=500$ and $1000$~GeV, 
for $\tan\beta =52$, $A_0=0$ and $\mu >0$. The colors on the various
regions and on the different contours are as in Fig.~\ref{fig:10plc}.
}
\label{fig:52plc}}

\section{Comparison of LC reach with Tevatron, LHC and $\Omega_{\tz_1}h^2$}

In this section, we present an overview of the reach of
a LC in comparison to the reach for sparticles that can be obtained by
the Fermilab Tevatron and the CERN LHC. In addition, we show
regions of relic neutralino dark matter density in accord with
the recent WMAP measurements.

Our first results are shown in Fig.~\ref{fig:10plcall} which shows
the same parameter space plane as in Fig.~\ref{fig:10plc}.
In this case, however, we plot the composite
reach plot of a $\sqrt{s}=0.5$ and 1~TeV LC for discovery of sparticles
assuming 100 fb$^{-1}$ of integrated luminosity. The LC reach plots now
consist of combined 1.) slepton pair reach via dileptons 
(we also show the 
kinematically accesible stau pair production contour, if it is
beyond the slepton  
reach), 
2.) the chargino pair 
reach via $1\ell +2j +\eslt$ events, with either the standard cuts
or the cuts specialized for searches in the HB/FP region, and 3.) the 
region of $\tz_1\tz_2 \to b\bar{b} +\eslt$. In addition, we superimpose 
on this plot the reach of the Fermilab Tevatron for SUSY via the
clean trilepton signal originating from $p\bar{p}\to\tw_1\tz_2 X
\to 3\ell +\eslt +X$, where $X$ denotes assorted hadronic debris.
The Tevatron reach was extended into the HB/FP region in Ref. \cite{bkt}; 
we show the optimistic reach assuming a $3\sigma$ signal with 
25 fb$^{-1}$ of integrated luminosity.
In addition, we show the reach of the CERN LHC as derived in
Ref. \cite{bbbkt}, assuming 100 fb$^{-1}$ of integrated luminosity.
We have also added to the plot the green region, which denotes
parameter space points with relic density $\Omega_{\tz_1}h^2<0.129$,
as required by the recent WMAP measurements.\footnote{The WMAP
allowed region including the lower bound would appear as a very narrow strip 
following the border of the green region.} For our relic density 
calculation\cite{bbb} we
have evaluated all relevant neutralino annihilation and co-annihilation
processes in the early universe using the CompHEP program.
We have implemented relativistic thermal averaging of the 
annihilation cross section times velocity, which is useful to get
the appropriate relic density in the vicinity of $s$-channel
poles (where the annihilating neutralinos may have substantial
velocities) using the formulae of Gondolo and Edsjo\cite{gondolo}.

The dark matter allowed region splits into three distinct regions
for $\tan\beta =10$. On the far left of the plot at low $m_0$
is the stau co-annihilation region, which blends into the
bulk annihilation region at low $m_{1/2}$ values. Note that the
bulk region is largely below the LEP2 $m_h=114.4$~GeV contour.
We also see that the stau co-annihilation region extends to
$m_{1/2}$ values as high as $\sim 900$~GeV.
For $\tan\beta =10$, we see that a $\sqrt{s}=0.5$ TeV
$e^+e^-$ collider should be able to scan much of the stau 
co-annihilation region, while a $\sqrt{s}=1$ TeV machine can 
cover it entirely (as can the LHC). Another region of relic density is
the small strip at constant $m_{1/2}\sim 100$~GeV, where neutralinos
can annihilate through the narrow $s$-channel pole from the light Higgs 
boson $h$. This region can be covered by all the colliders, including the
Fermilab Tevatron. Finally, adjacent to the REWSB excluded region
at large $m_0$ is shown the dark matter allowed region in the
HB/FP region, where the LSP has a significant higgsino component, which
facilitates neutralino annihilation to $WW$ and $ZZ$ pairs in the
early universe. The Fermilab Tevatron reach does not extend 
into this regime. The $\sqrt{s}=0.5$ TeV LC can 
explore the kinematically allowed portion of the 
lower HB/FP region (which is the region favored by
the fine-tuning analysis of Ref.\cite{fmm}) 
via standard cuts and the cuts specialized to the far HB/FP region.
The $\sqrt{s}=1$ TeV colider can explore {\it all} the HB/FP
region which is dark matter allowed, until $m_{1/2}$ becomes
greater than $\sim 900$~GeV. The portion of the HB/FP region
with $m_{1/2}>900$~GeV, while allowed by dark matter as well as  other
experimental constraints, becomes more difficult to reconcile with
fine-tuning considerations.
\FIGURE{\epsfig{file=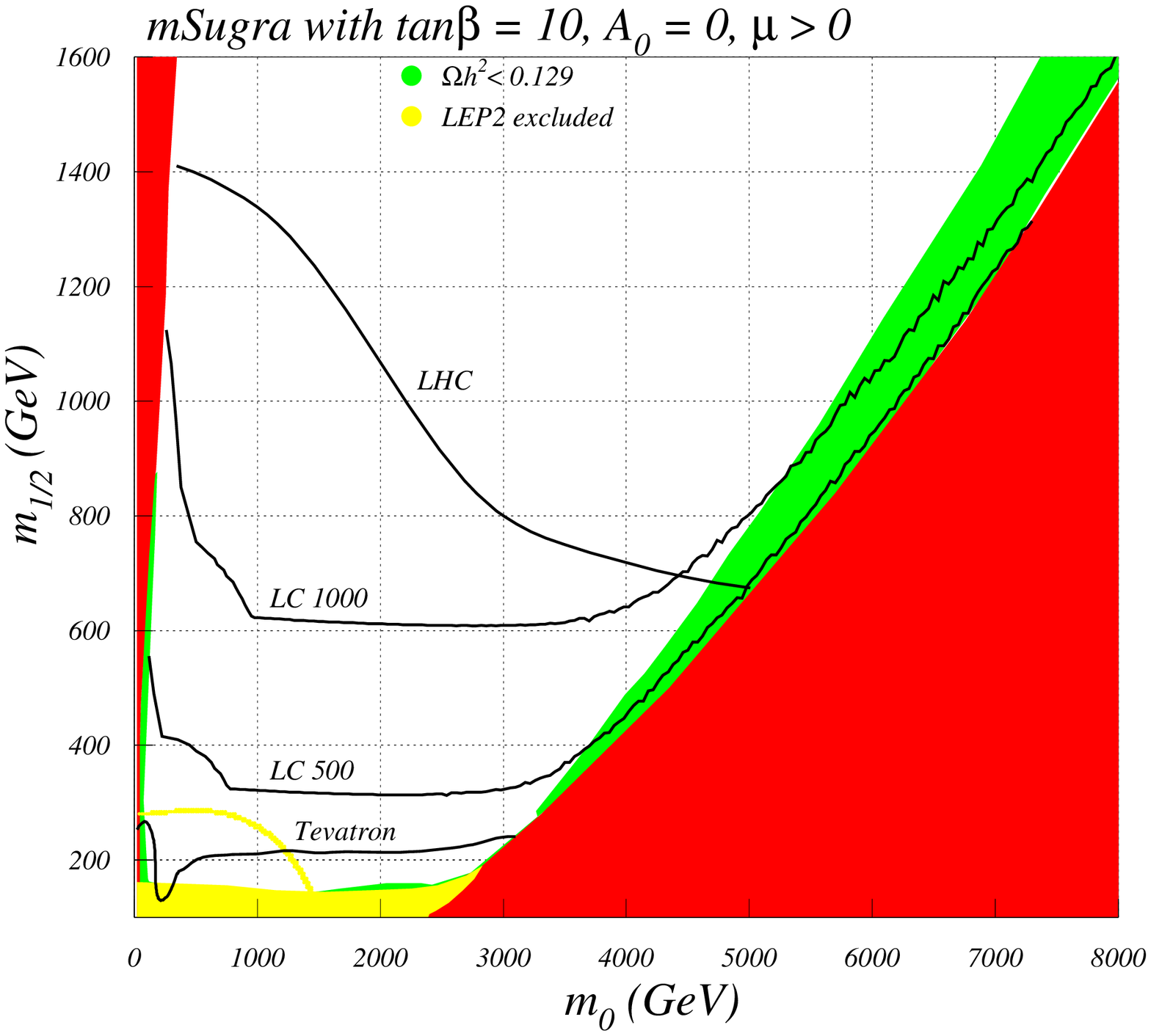,width=14cm} 
\vspace*{-1cm} 
\caption{Reach of a $\sqrt{s}=0.5$ and 1~TeV LC for sparticles
in the mSUGRA model for $\tan\beta =10$, $A_0=0$ and $\mu >0$.
We also show the reach of the Fermilab Tevatron assuming 10 fb$^{-1}$
of integrated luminosity (for isolated trileptons) and the reach of 
the CERN LHC assuming 100 fb$^{-1}$ of integrated luminosity.
Finally, the green shaded region shows points where the relic
density $\Omega_{\tz_1}h^2<0.129$ as dictated by WMAP.
}
\label{fig:10plcall}}
The new cuts proposed in Sec.~2 allow the LC SUSY search region to extend 
well beyond the reach of the CERN LHC, which extends only to
$m_{1/2}\sim 700$~GeV. The LHC reach is limited in the high $m_{1/2}$
part of the hyperbolic branch because sfermions and gluinos are too heavy
to be produced at an appreciable rate. Chargino and neutralino pairs
can still be produced at the LHC 
in the high $m_{1/2}$ part of the hyperbolic branch,
but the soft visible energy emanating from chargino and neutralino decay 
makes detection above background very difficult.
The high $m_{1/2}$ part of the hyperbolic branch yields a 
first example of a region of mSUGRA
model parameter space where {\it sparticles can be discovered at a LC, 
whereas 
the CERN LHC reach for sparticles has petered out.} Moreover, this
additional reach area comes in precisely at a very compelling dark matter
allowed region of the mSUGRA model.

In Fig.~\ref{fig:30plcall}, we show the same plot, except this time
for $\tan\beta =30$. Many features of the plot are 
qualitatively similar to the 
$\tan\beta =10$ case. 
In this case, the stau co-annihilation
corridor now extends up to $m_{1/2}$ values as high as
$1050$~GeV. The entire stau co-annihilation corridor can potentially be 
explored by a $\sqrt{s}=1$ TeV LC, but only if a stau pair search
is made in addition to the dilepton search. 
In this case, the Tevatron reach extends just to the tip
of the dark matter allowed HB/FP region. The LHC reach in the HB/FP 
region is again limited to $m_{1/2}<700$~GeV values, while
the $\sqrt{s}=0.5$ and especially the $\sqrt{s}=1$ TeV LC
can explore much of the HB/FP region, even for 
$m_{1/2}$ values far in excess of 700~GeV.
\FIGURE{\epsfig{file=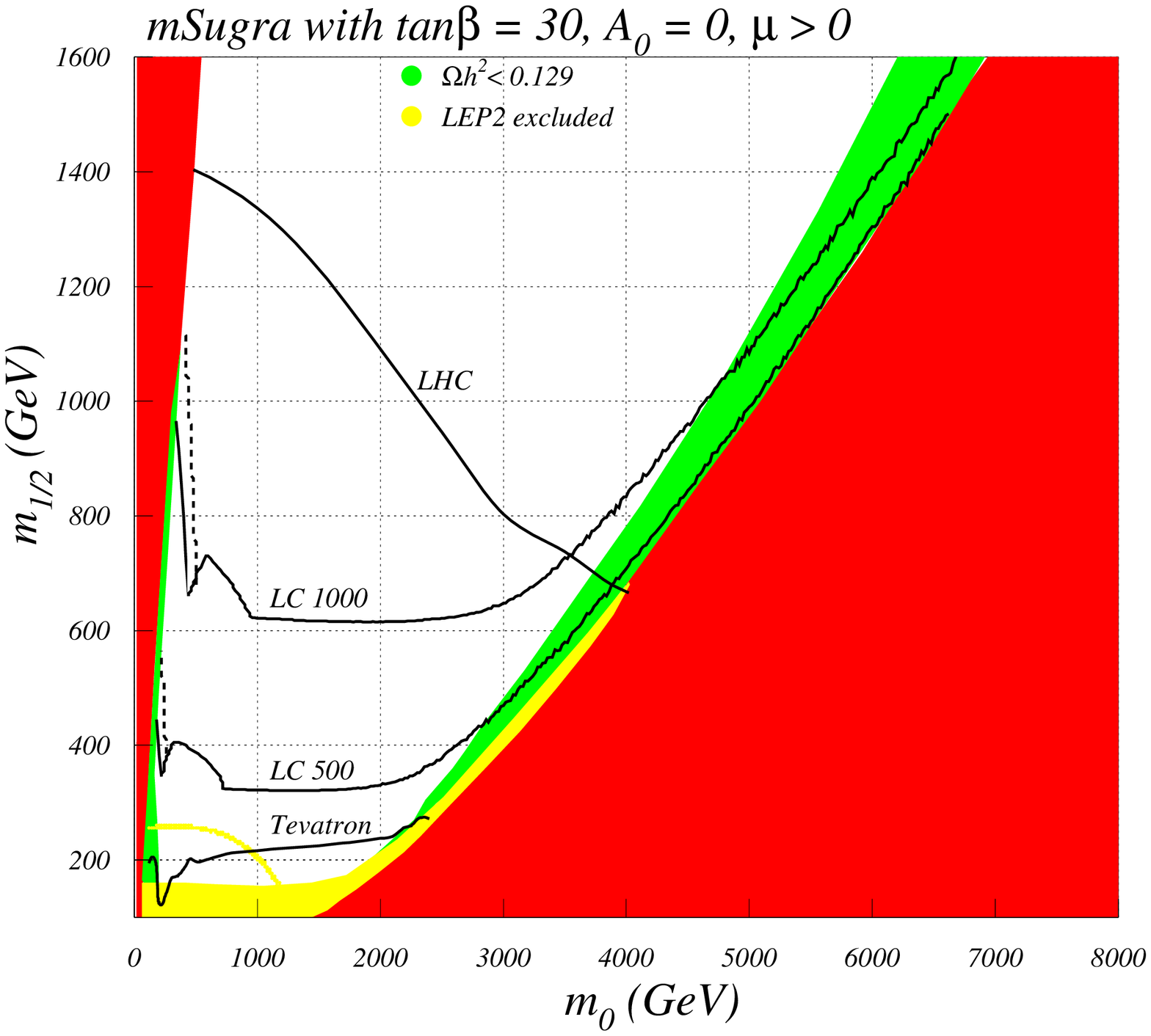,width=14cm} 
\vspace*{-1cm} 
\caption{Reach of a $\sqrt{s}=0.5$ and 1~TeV LC for sparticles
in the mSUGRA model for $\tan\beta =30$, $A_0=0$ and $\mu >0$.
We also show the reach of the Fermilab Tevatron assuming 10 fb$^{-1}$
of integrated luminosity (for isolated trileptons) and the reach of 
the CERN LHC assuming 100 fb$^{-1}$ of integrated luminosity.
Finally, the green shaded region shows points where the relic
density $\Omega_{\tz_1}h^2<0.129$ as dictated by WMAP.
We denote the kinematic limit for stau pair production at LCs
by a dashed black contour.
}
\label{fig:30plcall}}

The dark matter relic density is qualitatively different for
the case of $\tan\beta =45$, $\mu <0$ shown in Fig.~\ref{fig:45mlcall}.
Here, a large new dark matter allowed region has emerged, namely the
$A$-annihilation funnel which is characteristic of the mSUGRA model at very 
large $\tan\beta$. As $\tan\beta$ increases, the derived value
of $m_A$ decreases, until a region where $m_A\simeq 2m_{\tz_1}$
arises, where neutralinos can efficiently annihilate
through the very broad $A$ and also the $H$ $s$-channel
resonances. It can be seen from the figure that the 
$A$-annihilation funnel region extends well beyond the reach of both 
the $\sqrt{s}=0.5$ and 1~TeV LC. In addition, the stau
co-annihilation strip rises to $m_{1/2}$ values that are also beyond the reach
of a 1~TeV LC.  The CERN LHC can explore
essentially all of the $A$-annihilation funnel for this
particular value of $\tan\beta$ and sign of $\mu$.
Also, in this case, the $\sqrt{s}=0.5$ TeV LC can explore
the HB/FP region only up to $m_{1/2}\sim 600$~GeV where the
$m_{\tw_1}=250$~GeV contour intersects the excluded region.
The 1~TeV LC has a reach that extends again well beyond the limit of
the LHC reach in the HB/FP region.
\FIGURE{\epsfig{file=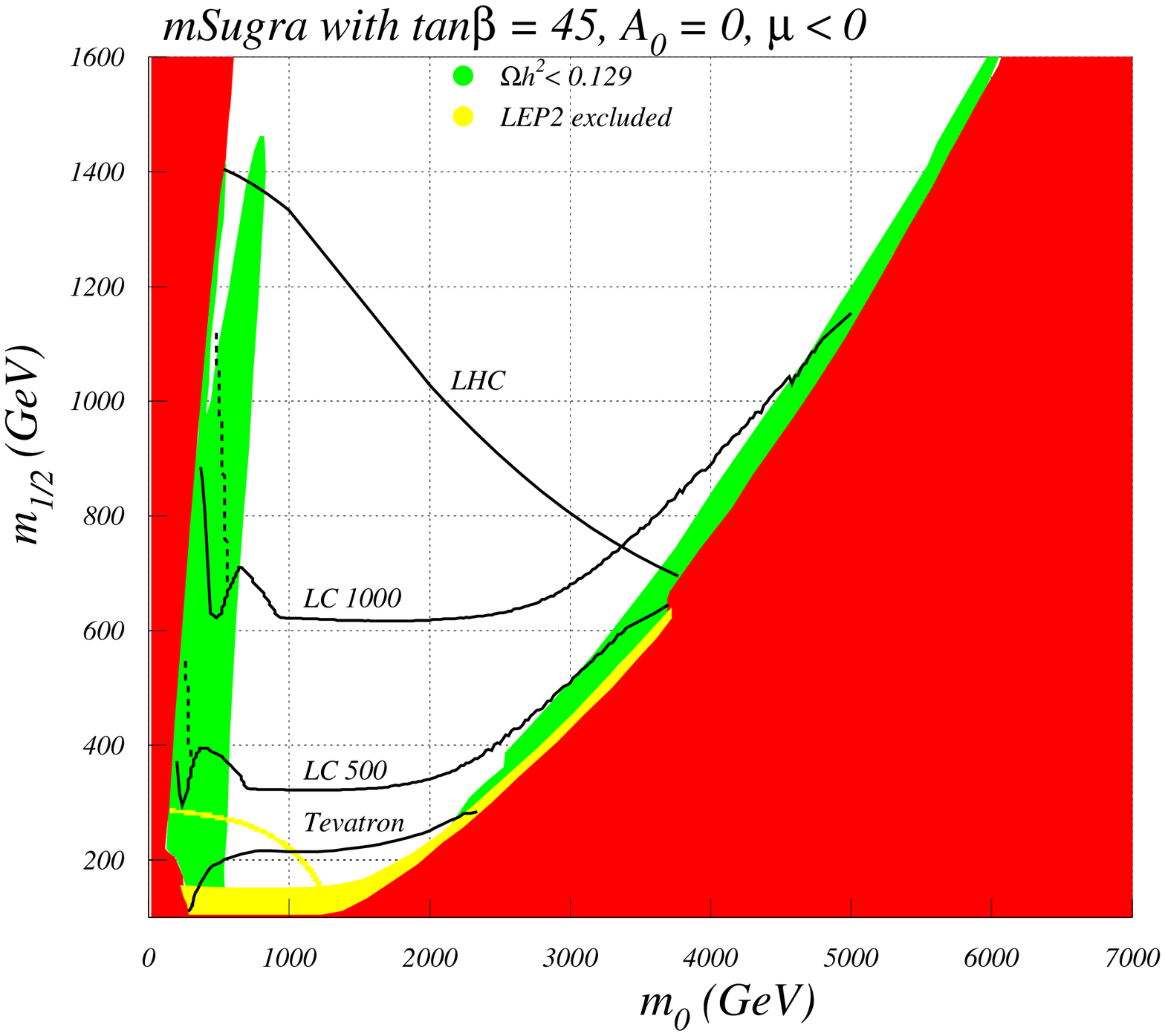,width=14cm} 
\vspace*{-1cm} 
\caption{Reach of a $\sqrt{s}=0.5$ and 1~TeV LC for sparticles
in the mSUGRA model for $\tan\beta =45$, $A_0=0$ and $\mu <0$.
We also show the reach of the Fermilab Tevatron assuming 10 fb$^{-1}$
of integrated luminosity (for isolated trileptons) and the reach of 
the CERN LHC assuming 100 fb$^{-1}$ of integrated luminosity.
Finally, the green shaded region shows points where the relic
density $\Omega_{\tz_1}h^2<0.129$ as dictated by WMAP.
We denote the kinematic limit for stau pair production at LCs
by a dashed black contour.}
\label{fig:45mlcall}}

In Fig.~\ref{fig:52plcall}, we show the $\tan\beta =52$ mSUGRA plane
for $\mu >0$. In this case, the effect of the $A$-annihilation funnel is just 
beginning to enter the $m_0\ vs.\ m_{1/2}$ plane from the left, so that
points along the low $m_0$ forbidden region have a low relic density because
neutralinos can annihilate via stau coannihilation, via $t$-channel
slepton (mainly stau) exchange (low $m_{1/2}$) {\it and} partly due to
annihilation through the $s$-channel $A$ resonance. In this case, the $A$
resonance corridor is actually off the plot, but since the $A$ width is 
so large ($\Gamma_A\sim 25$~GeV for $m_{1/2}\sim 600$~GeV), the
value of $2m_{\tz_1}$ can be a few partial widths away from resonance and
still give significant contributions to the neutralino annihilation
rate.
For this large a $\tan\beta$ value, the stau co-annihilation strip
reaches $m_{1/2}$ values far beyond the reach of LCs or even the LHC.
In the HB/FP region, the new cuts presented in Sec.~2 again give the LCs
a reach well beyond the LHC for $m_{1/2}>700$~GeV.
\FIGURE{\epsfig{file=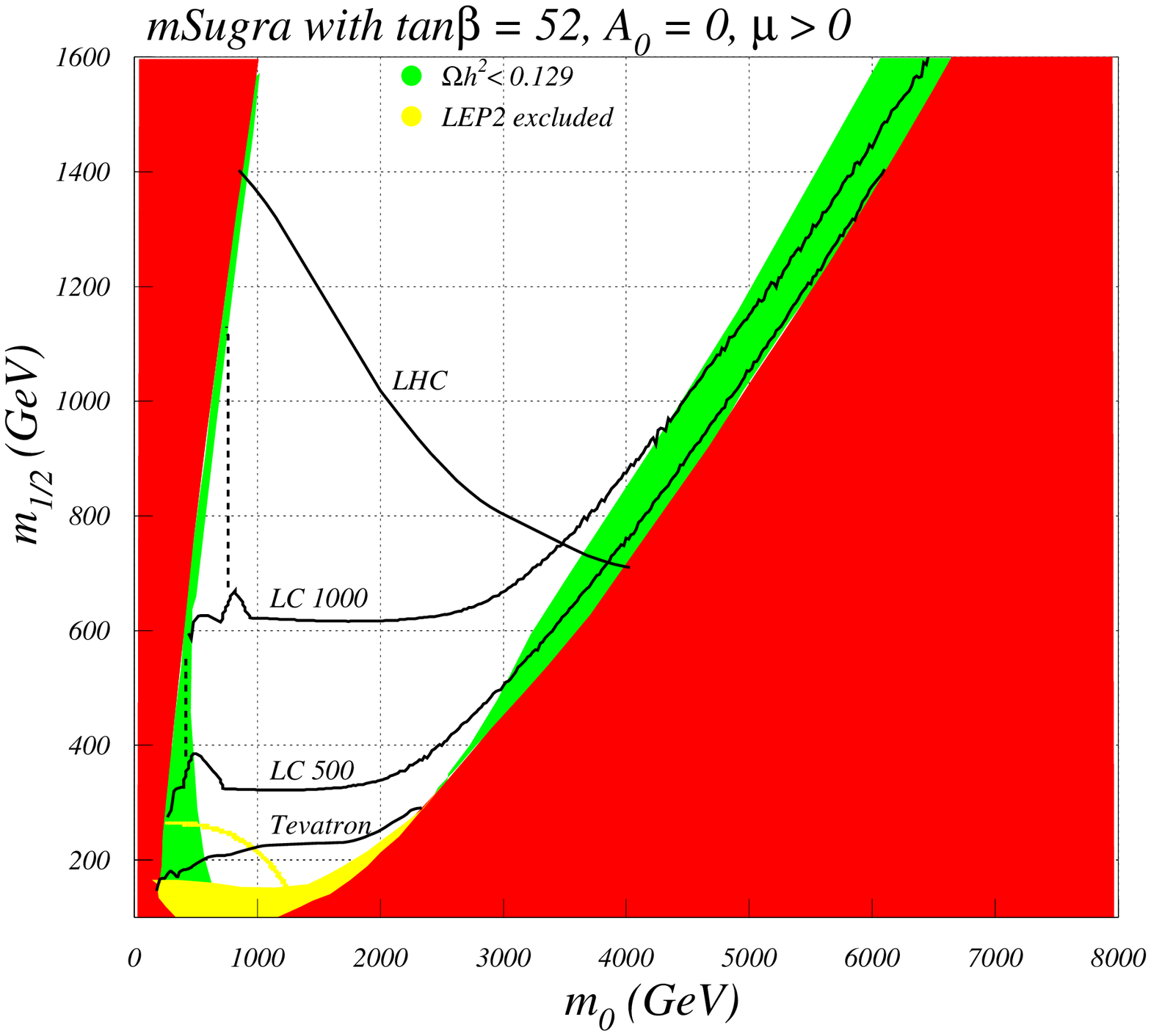,width=14cm} 
\vspace*{-1cm} 
\caption{Reach of a $\sqrt{s}=0.5$ and 1~TeV LC for sparticles
in the mSUGRA model for $\tan\beta =52$, $A_0=0$ and $\mu >0$.
We also show the reach of the Fermilab Tevatron assuming 10 fb$^{-1}$
of integrated luminosity (for isolated trileptons) and the reach of 
the CERN LHC assuming 100 fb$^{-1}$ of integrated luminosity.
Finally, the green shaded region shows points where the relic
density $\Omega_{\tz_1}h^2<0.129$ as dictated by WMAP.
We denote the kinematic limit for stau pair production at LCs
by a dashed black contour.}
\label{fig:52plcall}}

\section{Determination of Model Parameters in the HB/FP Region}

Once a signal for supersymmetry is established at a LC, then
the next task will be to scrutinize the signal to 
elucidate production and decay processes, extract sparticle  
masses, spins and other quantum numbers, and ultimately
to determine parameters of the underlying model.
Many groups have examined different case 
studies\cite{lcstudies}. 
In this section, we will examine a
case study in the low $m_{1/2}$ part of the HB/FP region in an
attempt to extract the underlying parameters of the MSSM, which
may in turn point to nature actually being described by parameters
in the HB/FP region. 

Toward this end, we consider Case 2, with the mSUGRA parameters set
given by
$$m_0,\ m_{1/2},\ A_0,\ \tan\beta ,\ sign(\mu ) = 
2500 \ {\rm GeV}, 300 \ {\rm GeV}, 0, 30, +1 .$$ 
Sample sparticle masses and parameters
are given in Table~\ref{tab:cs2}. For these parameter choices, 
$|\mu |< M_2$, so that the light chargino and lightest neutralino have
significant higgsino components. The chargino mass
$m_{\tw_1}=113.1$~GeV, and is just beyond the reach of LEP2. The LSP mass is
$m_{\tz_1}=85.6$~GeV, so that the mass gap 
$m_{\tw_1}-m_{\tz_1}=27.5$~GeV. The $\tw_1$ decays via 3-body modes
into $\tz_1 f\bar{f}'$, where $f$ and $f'$ are SM fermions. 
The decays are dominated by the $W$ boson exchange graphs, so that decays
$\tw_1\to\tz_1 f\bar{f}'$ have similar branching fractions to
$W\to f\bar{f}'$ decays.

We begin by generating $e^+e^-\to$ all SUSY particles for the signal,
and generate SM backgrounds using all ISAJET SM processes.
We first require all events to pass the standard chargino pair cuts for
$1\ell +2j +\eslt$ events as detailed at the beginning of Sec.~2.
Next, following case study 4 of Ref. \cite{bmt}, we require
missing mass $\mslash >240$~GeV. The resulting signal and also background
events are plotted in Fig.~\ref{fig:ejjmjj} in the $E(jj)\ vs.\ m(jj)$ plane.
SUSY and Higgs boson
\FIGURE{\epsfig{file=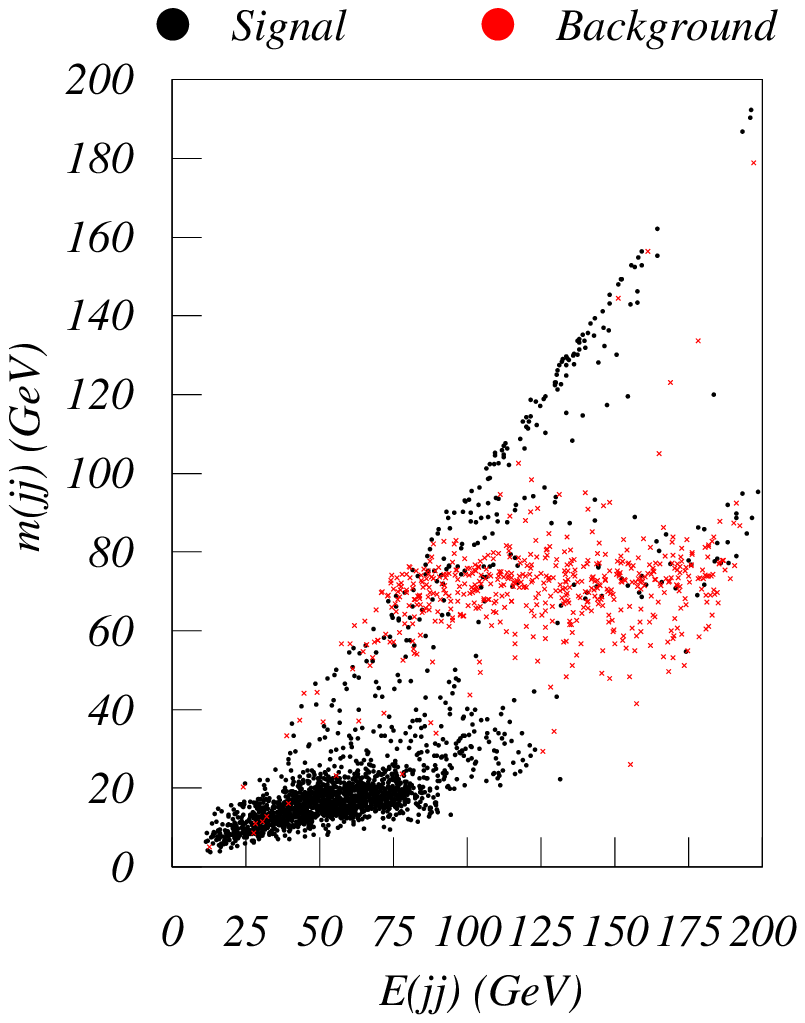,width=10cm} 
\vspace*{-1.cm}
\caption{Scatter plot of SUSY signal events (black dots) 
and SM background (red x's) 
after standard cuts plus $\mslash >240$~GeV cuts, in the
$E_{jj}\ vs.\ m(jj)$ plane. Chargino pair events occupy
the low $m(jj)$ region.
}
\label{fig:ejjmjj}}
events are denoted by black dots, while SM background events are
denoted by red crosses. The chargino pair events populate the cluster
at low $m(jj)$, since the dijet mass from chargino decay is bounded by
the $\tw_1 -\tz_1$ mass difference. If the chargino decays  via
$\tw_1 \to W\tz_1 \to q\bar{q}'\tz_1$, we would expect that 
the $E(jj)$ distribution 
would have well-defined upper and lower endpoints that depend only
on $m_{\tw_1}$ and $m_{\tz_1}$ (and, of course $M_W$), 
as in case study 1 of Ref. \cite{jlc1}.

In the HB/FP region, the $\tw_1-\tz_1$ mass gap is small, and the 
decay to on-shell $W$ is kinematically inaccessible. We can, however,
adapt this strategy by forcing ``two-body kinematics'' on these 
events by first selecting events in narrow bins in $m(jj)$ and studying
separately their $E(jj)$ distributions. This is done 
in Fig.~\ref{fig:ejjbins}, where we show the $E(jj)$ distribution for
$m(jj)$ bins of width 4~GeV, centered at 8, 12, 16 and 20~GeV,
corresponding to an integrated luminosity of 100~fb$^{-1}$. This 
is the ``data''.
The energy of the dijet cluster is bounded by
\be
\gamma (E_{jj}^*-\beta p_{jj}^*)\le E(jj)\le\gamma (E_{jj}^*+\beta p_{jj}^*),
\label{eminemax}
\ee  
where $E_{jj}^*=(m_{\tw_1}^2+m^2(jj)-m_{\tz_1}^2)/2m_{\tw_1}$, 
$p_{jj}^*=\sqrt{E_{jj}^{*2}-m^2(jj)}$, $\gamma =E_{\tw_1}/m_{\tw_1}$, 
$\beta =p_{jj}^*/ E_{jj}^*$ and 
$E_{\tw_1}=\sqrt{s}/2$, up to 
energy mismeasurement errors, jet clustering, particle losses,
bremsstrahlung and finite width bins in $m(jj)$. 
The corresponding ``theoretical predictions'' shown by the smooth curve
are obtained by generating a much larger sample of the same events and 
fitting this larger sample (corresponding to an integrated luminosity of
600~fb$^{-1}$) to the function,
%
{\small
\be
F(E,m_{\tw_1},m_{\tz_1};A,B,C,D)=N\left\{1+\exp\left [\frac{E_{min}+A-E}{B\sigma_{E_{min}}}\right ]\right\}^{-1}\left\{1+\exp\left [\frac{-E_{max}+C+E}{D\sigma_{E_{max}}}\right ]\right\}^{-1},
\label{fitfun}
\ee
}
where $E_{min}$ and $E_{max}$ are calculated for each bin in $m(jj)$ taking 
the central $m(jj)$ value in that bin, and input values for $m_{\tw_1}$ and 
$m_{\tz_1}$; $\sigma_{E_{min}}$ 
($\sigma_{E_{max}}$) is the absolute value of the difference between 
$E_{min}$ ($E_{max}$) at the highest $m(jj)$ value in each $m(jj)$ bin and 
$E_{min}$ ($E_{max}$) for $m(jj)$ at the center of this bin. The small
contribution from the SM background has also been included.
\FIGURE{\epsfig{file=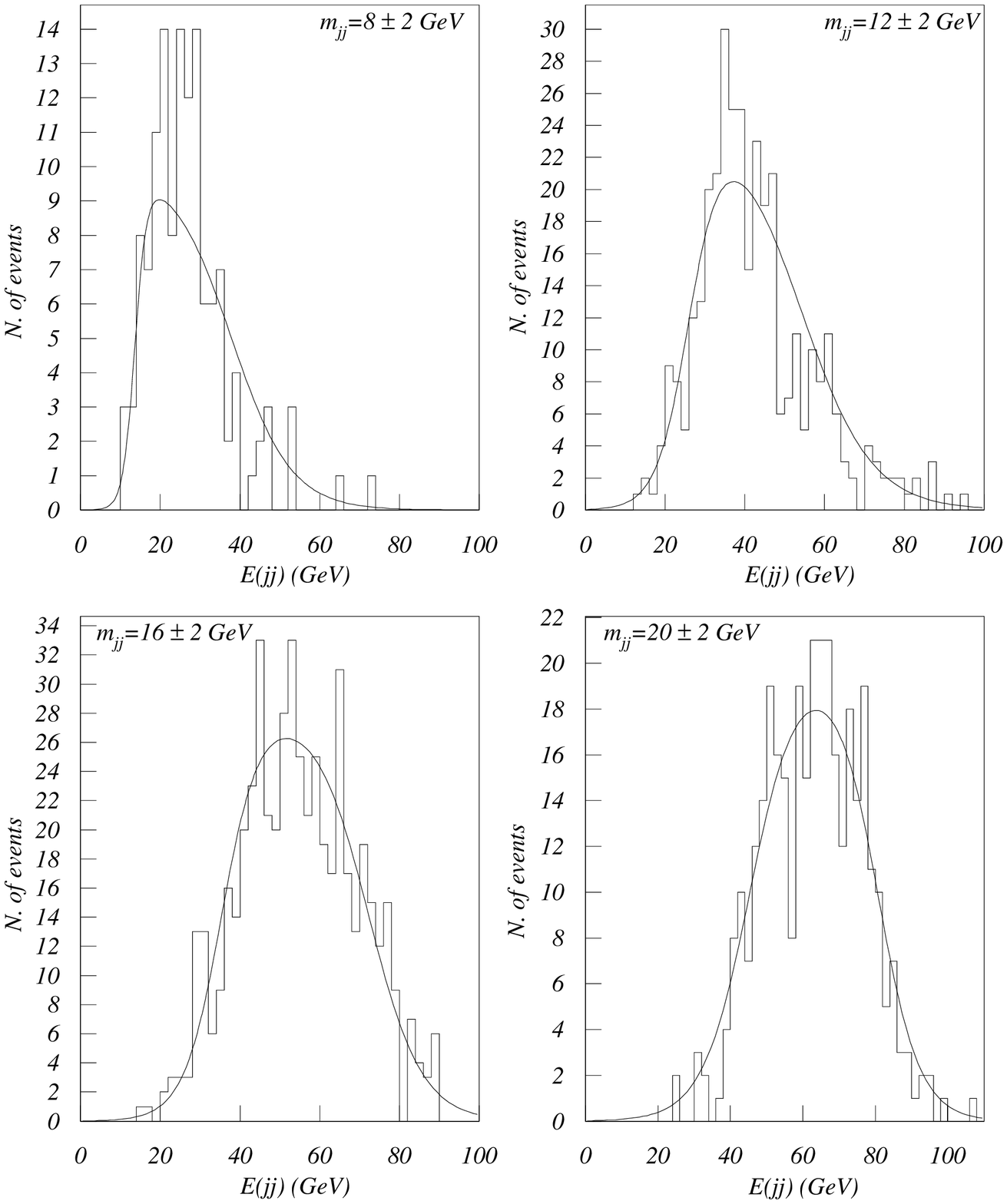,width=14.5cm} 
\vspace*{-1.cm}
\caption{Distribution of $E_{jj}$ for $\ell +2-jet$ events after
standard cuts together with $\mslash >240$~GeV, with events restricted to
narrow bins of $m(jj)$. The histograms show these distributions for
the synthetic data sample while the solid line shows the corresponding
theoretical expectation obtained as described in the text.}
\label{fig:ejjbins}}
The parameters  $A$, $B$, $C$ and $D$ are separately 
determined for each bin 
in $m(jj)$ and serve to fit the shapes of the corresponding
distributions, while 
$N$ is adjusted to give the normalization 
near the maximum of $E_{jj}$ distribution
corresponding to an 
integrated luminosity of 100~fb$^{-1}$.

Next, we proceed to perform a $\chi^2$ fit to obtain $m_{\tw_1}$ and
$m_{\tz_1}$ from our synthetic data sample, using the fitted function
(\ref{fitfun}) for the theoretical prediction \footnote{Of course, the 
parameters $E_{min}$, $\sigma_{E_{min}}$, $E_{max}$ and 
$\sigma_{E_{max}}$ depend on $m_{\tw_1}$ and $m_{\tz_1}$ via (\ref{eminemax}) 
and the equations following that. We assume that the parameters $A, B,
C$ and $D$ do not change.}
for chargino and neutralino
masses close to those for Case 2. In other words, for a grid of points 
in the ($m_{\tw_1}, m_{\tz_1}$) plane, we evaluate,
{\small
\be
\chi^2(m_{\tw_1},m_{\tz_1})
=\sum_{bins}\sum_{E}\left (\frac{F(E,m_{\tw_1}(inp), m_{\tz_1}(inp))
-F(E,m_{\tw_1},m_{\tz_1})}{\sqrt{F(E,m_{\tw_1},m_{\tz_1})}}\right )^2
\ee
}
where $\sum_{bins}$ means that we sum over all four bins in $m(jj)$, and 
$\sum_{E}$ denotes the summation over all bins in $E_{jj}$ and find 
the values of chargino and neutralino masses for which this quantity is
minimized.  
These best fit values, together with the regions where $\Delta \chi^2 \le
2.3$~(68.3\% CL) and $\le 4.6$ (90\% CL)  
are shown in Fig.~\ref{fig:mz1mw1}. We see that it is possible to
determine $m_{\tw_1}$ and $m_{\tz_1}$ at approximately the 10\% level. 
\FIGURE{\epsfig{file=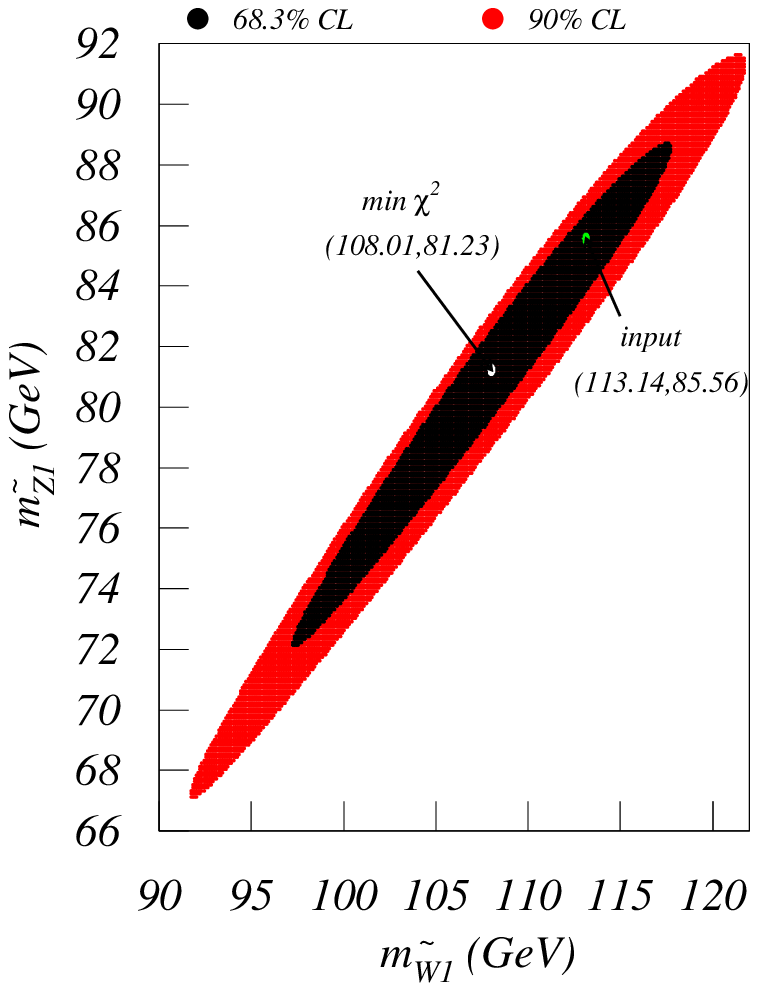,width=9cm} 
\vspace*{-1.cm}
\caption{Fits to $m_{\tw_1}$ and $m_{\tz_1}$ and the associated error 
ellipses for Case 2 in the text.
}
\label{fig:mz1mw1}}

Having determined the values of $m_{\tw_1}$ and $m_{\tz_1}$, 
the next step is to examine what we can say about 
the underlying MSSM parameters $\mu$, $M_2$ and $\tan\beta$ that enter
the chargino mass matrix. To determine three unknowns, we need to 
experimentally determine one more quantity which we take to be 
the 
cross section for $1\ell+ 2j+\eslt$ events from
chargino pair production. Almost all the signal arises from
chargino pair production  if we
require $m(jj)<25$~GeV and $E(jj)<100$~GeV as in 
Fig.~\ref{fig:ejjmjj}. For 100 fb$^{-1}$ of integrated luminosity,
we find 1649 events, which translates to a measurement of
$\sigma (e^+e^-\to\tw_1^+\tw_1^- )=16.5\pm 0.4$~fb (after all the cuts), 
or 2.5\% statistical error.
Other systematic errors will be present, although these may be 
controllable by precision measurement of many SM processes.
We perform a fit to the MSSM parameters
using the values of 
$m_{\tw_1}$, $m_{\tz_1}$ and $\sigma (\tw_1^+\tw_1^- )$ as determined
above. 
We scan over MSSM
model parameters, using 1-loop corrected mass relations for
$m_{\tw_1}$ and $m_{\tz_1}$ as given by ISAJET 7.69. In
Fig.~\ref{fig:m2mutanb},
we show the regions of
{\it a}) the $\mu\ vs.\ M_2$ plane, {\it b}) the 
$\mu\ vs.\ \tan\beta$ plane and {\it c}) the $M_2\ vs.\ \tan\beta$ plane
that are allowed at the 68.3\% and 90\% CL. In each case, we have 
held the parameter not shown in the plane fixed at its input value.
The result in frame {\it a}) clearly shows that indeed $|\mu |<<M_2$,
providing strong support that the model parameters 
lie in the HB/FP region, and
that the LSP has a significant higgsino component, enhancing the
neutralino pair annihilation in the early universe.\footnote{That
$|\mu|$ is small can presumably also be determined by studying chargino
production using polarized beams. Note that our determination does not
require this capability.} While $\mu$ and $M_2$ can be well determined
(at least for this case study), it is also evident from the figure that
a precise determination of $\tan\beta$ is not possible in this case.
This may not be so surprising, since in the HB/FP region, SUSY scalar
masses that depend on Yukawa couplings and hence $\tan\beta$ are so
heavy that they essentially decouple from observable physics, and the
region is relatively invariant under changes in $\tan\beta$.

\TABLE{
\begin{tabular}{lc}
\hline
parameter & value (GeV) \\
\hline
$M_2$ & 236.5 \\
$M_1$ & 122.0 \\
$\mu$ & 121.6 \\
$m_{\tg}$ & 833.2 \\
$m_{\tu_L}$ & 2548.1 \\
$m_{\te_L}$ & 2503.9 \\
$m_{\tw_1}$ & 113.1 \\
$m_{\tw_2}$ & 274.8 \\
$m_{\tz_1}$ & 85.6 \\
$m_{\tz_2}$ & 135.0 \\ 
$m_{\tz_3}$ & 142.2 \\
$m_{\tz_4}$ & 281.5 \\
$m_A$ & 2129.4 \\
$m_h$ & 118.8 \\
$\Omega_{\tz_1}h^2$& 0.0423\\
$\scriptsize BF(b\to s\gamma)$ & $3.84\times 10^{-4}  $\\
$\Delta a_\mu                $ & $2.3 \times  10^{-10}$\\ 
\hline
\label{tab:cs2}
\end{tabular}
\vspace*{-0.4cm}
\caption{Masses and parameters in~GeV units for Case 2 
for $m_0,\ m_{1/2},\ A_0,\ \tan\beta ,\ sign\mu =$
2500~GeV, 300~GeV, 0, 30, +1 in the mSUGRA model.
The spectra is obtained using ISAJET v7.69.}
}
\FIGURE{\epsfig{file=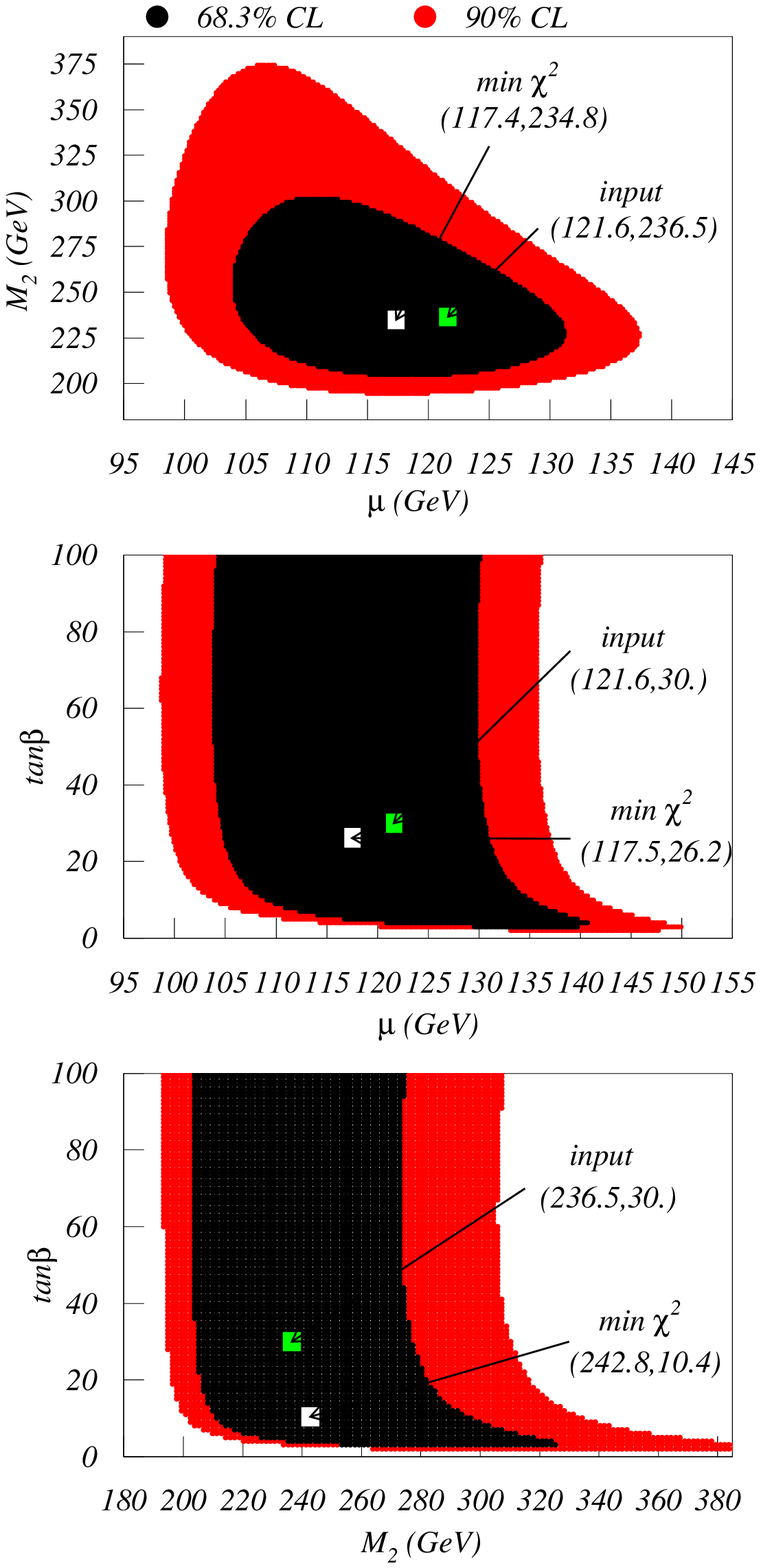,width=11cm} 
\vspace*{-1.cm}
\caption{Fits to $\mu$, $M_2$ and $\tan\beta$
from the measured values of $m_{\tw_1}$, $m_{\tz_1}$ and
$\sigma (\tw_1^+\tw_1^- )$, and the associated 68.3\% and 90\% CL
regions for Case 2. The green squares denote the fitted values
while the white squares show the corresponding input values of the parameters.
}
\label{fig:m2mutanb}}

\section{Summary and conclusions}

The recent constraint on the relic density of 
neutralinos obtained from WMAP measurements, together with earlier
determinations of 
$BF(b\to s\gamma )$ and $(g-2)_\mu$ select out regions
of parameter space of the mSUGRA model. In the stau co-annihilation
region, the $H,A$-annihilation funnel and in the HB/FP regions, 
very high values of $m_0$ and $m_{1/2}$ consistent
with all constraints are possible: moreover, the so-called bulk region where 
sparticles are light is disfavored.
These considerations motivated us to re-assess the
reach of various collider and non-accelerator search experiments for
supersymmetry. In this paper, we re-evaluate the reach of a 
$\sqrt{s}=0.5$ and 1~TeV linear $e^+e^-$ collider for SUSY in the
context of the mSUGRA model, examining for the first time the reach
in the HB/FP region. We find that a $\sqrt{s}=1$ TeV LC can explore
most of the stau co-annihilation region if $\tan\beta \alt 30$,
although along with a dilepton search,  a ditau search will 
also be needed. The $H,A$-annihilation funnel typically extends beyond 
the maximum reach of a LC. In the HB/FP region, chargino pairs may be
kinematically accessible to a LC, but the energy release in chargino
pairs can be small, reducing detection efficiency. Nonetheless,
LCs should be able to probe much of the lower HB/FP region with 
standard chargino searches. In the upper HB/FP region, new cuts are proposed
to allow signals from 
much of the small $\tw_1 -\tz_1$ mass gap region to be
observable above SM backgrounds. 
In this region, the reach of even a 500~GeV LC can
exceed that of the CERN LHC! This is all the more important in that it 
occurs in a region of model parameter space which is allowed by all
constraints, including those imposed by WMAP.

One should also stress that the LCs reach is also complimentary
to reach of direct dark matter search experiments (DDMS)
even though both kinds of experiments 
similarly cover much of the HB/FP region~\cite{ddmsearch}.
The complementarity of a LC occurs for the region 
which is very close to the no REWSB border.
In this region, the neutralino relic density is so low that DDMS experiments
are  not able to cover this part of the parameter space
even though the higgsino component 
of neutralino is significant. 
This region can be probed by experiments at a LC.

If a supersymmetric signal is found, then the next obvious step will be
to determine the
underlying MSSM parameters. We have performed a case study in the 
low $m_{1/2}$ part of the hyperbolic branch.
In this region, we show that a measurement of $m_{\tw_1}$
and $m_{\tz_1}$ is possible at the 10\% level. A measurement of the 
total chargino pair cross section to 2.5\% allows a 
determination of MSSM parameters $M_2$ and $\mu$, although 
$\tan\beta$ is more difficult to pin down. The resulting determination
of $M_2$ and $\mu$
would point to a model with higgsino-like charginos and 
neutralinos. Together with absence (or low levels)
of squark signals at the LHC,
and the agreement of the chargino cross section with the expected
$s$-channel contribution (pointing to heavy sneutrinos) these measurements
would be indicative of an mSUGRA-type model
in the HB/FP region.  In  case parameters are in the upper
part of the hyperbolic branch, LC event characteristics may be 
sufficient at least to establish the production of massive particles,
with associated decay products that are quite soft. An examination  
of how one would obtain information about the underlying scenario
would be worthy of exploration. 

\acknowledgments
 
This research was supported in part by the U.S. Department of Energy
under contracts number DE-FG02-97ER41022 and DE-FG03-94ER40833.
	
%

\end{document}